\documentclass[english,11pt,oneside]{article}
\pdfoutput=1
\usepackage{amsmath}
\usepackage{amsfonts}
\usepackage{amssymb}
\usepackage{graphicx, rotating}
\usepackage{epstopdf}
\usepackage{epsfig}
\usepackage{latexsym}
\usepackage{multirow}
\usepackage{color}
\usepackage{slashed}
\usepackage{pifont}
\usepackage[top=2.8 cm, bottom=2.8 cm, left=3 cm, right=3 cm]{geometry}
\usepackage[compat=1.0.0]{tikz-feynman}
\usepackage[bookmarks]{hyperref}
\usepackage{fancyhdr}
\usepackage[normalem]{ulem}

\usepackage[nottoc]{tocbibind}
\usepackage{cite}

\newcommand{\ba}{\begin{eqnarray}}
\newcommand{\ea}{\end{eqnarray}}
\newcommand{\no}{\nonumber}

\usepackage{hyperref}
\hypersetup{
 colorlinks = true,
 linkcolor  = red,
 citecolor={blue}
}

\usepackage{bbold}
\newcommand{\newc}{\newcommand}

\newc{\lcal}{\int {\cal L}dt}
\newc{\ie}{{\it i.e.}}          
\newc{\etal}{{\it et al.}}
\newc{\eg}{{\it e.g.}}          
\newc{\kev}{\hbox{\rm\,keV}}            
\newc{\mev}{\hbox{\rm\,MeV}}            
\newc{\gev}{\hbox{\rm\,GeV}}            
\newc{\tev}{\hbox{\rm\,TeV}}
\newc{\xpb}{\hbox{\rm\, pb}}
\newc{\xfb}{\hbox{\rm\, fb}}

\def\bea{\begin{eqnarray}}
\def\eea{\end{eqnarray}}
\def\beq{\begin{equation}}
\def\eeq{\end{equation}}

\def\lsim{\mathrel{\rlap{\lower3pt\hbox{\hskip0pt$\sim$}}
   \raise1pt\hbox{$<$}}}         
\def\gsim{\mathrel{\rlap{\lower4pt\hbox{\hskip1pt$\sim$}}
   \raise1pt\hbox{$>$}}}         

\begin{document}

\title{The CKM Phase and $\bar\theta$ in Nelson-Barr Models
}
\author{Alessandro Valenti$^{1\,2}$~\footnote{\href{mailto:alessandro.valenti@pd.infn.it}{\color{black}{alessandro.valenti@pd.infn.it}}}~~~~Luca Vecchi$^2$~\footnote{\href{mailto:luca.vecchi@pd.infn.it}{\color{black}{luca.vecchi@pd.infn.it}}}\\
{$^1$\small\emph{Dipartimento di Fisica e Astronomia ``G. Galilei", Universit\'a di Padova, Italy}}\\
{$^2$\small\emph{Istituto Nazionale di Fisica Nucleare (INFN), Sezione di Padova, Italy}}}
\date{}
\maketitle

\begin{abstract}

We analyze the Nelson-Barr approach to the Strong CP Problem. We derive the necessary conditions in order to simultaneously reproduce the CKM phase and the quark masses. Then we quantify the irreducible contributions to the QCD topological angle, namely the corrections arising from loops of the colored fermion mediators that characterize these models. Corrections analytic in the couplings first arise at 3-loop order and are safely below current bounds; non-analytic effects are 2-loop order and decouple as the mediators exceed a few TeV. We discuss collider, electroweak, and flavor bounds and argue that most of the parameter space above the TeV scale is still allowed in models with down-type mediators, whereas other scenarios are more severely constrained. With two or more families of mediators the dominant experimental bound is due to the neutron electric dipole moment.

\end{abstract}

\newpage

{
	\hypersetup{linkcolor=black}
	\tableofcontents
}

\section{Introduction}
\label{sec:intro}

In the renormalizable version of the Standard Model (SM), CP violation is controlled by two parameters: the CKM phase and the QCD theta angle.~\footnote{To be precise, there are two topological angles in the SM. However, because the electroweak theta angle can be entirely hidden in the abelian sector (and ultimately ends up being the topological angle of QED), it is practically irrelevant in collider experiments and can be ignored. It would be observable, for example, in the presence of topological defects. On the contrary, the QCD theta angle is both un-removable, since the associated gauge theory is intrinsically vector-like, as well as physical, because of the topological character of non-abelian gauge theories.} The former is parametrized by the following re-scaling invariant combination of the Yukawa matrices $Y^{\rm SM}_u, Y^{\rm SM}_d$
\ba\label{Jarlskog}
2J
=\frac{{\rm Im}~{\rm det}[Y^{\rm SM}_u{Y^{\rm SM}_u}^\dagger,Y^{\rm SM}_d{Y^{\rm SM}_d}^\dagger]}{\prod_{i,j\neq i}\left((\widehat Y_u^{\rm SM})^2_i-(\widehat Y_u^{\rm SM})^2_j\right)\prod_{i,j\neq i}\left((\widehat Y_d^{\rm SM})^2_i-(\widehat Y_d^{\rm SM})^2_j\right)}
\ea
where $(\widehat Y_u^{\rm SM})^2_i, (\widehat Y_d^{\rm SM})^2_i$ are the real eigenvalues of the hermitian matrices $Y^{\rm SM}_u{Y_u^{\rm SM}}^\dagger,Y^{\rm SM}_d{Y^{\rm SM}_d}^\dagger$. The numerical value \cite{PDG}
\ba\label{Jmeasured}
J\simeq3.0\times10^{-5}
\ea
is measured in a multitude of flavor-violating observables. Adopting the Wolfenstein parameterization of the CKM matrix one gets $J= A^2\lambda_C^6\eta(1+{\cal O}(\lambda_C^2))$, and eq. \eqref{Jmeasured} is understood to be the result of small mixing angles controlled by the Cabibbo angle $\lambda_C\simeq0.23$ and a CP-odd parameter of order one, i.e. $\eta\sim0.37$.

The QCD theta angle is instead parametrized by the re-scaling invariant parameter: 
\ba\label{bartheta}
\bar\theta=\theta-{\rm Arg}\left({\rm det}[Y^{\rm SM}_u]\,{\rm det}[Y^{\rm SM}_d]\right).
\ea
An important observable sensitive to $\bar\theta$ is the neutron electric dipole moment, $d_n$. It is very hard to make concrete predictions for $d_n$ due to the non-perturbative nature of low-energy QCD. The best one can do analytically is to use naive dimensional analysis. Moving $\bar\theta$ in the phase of the quark masses, and working at leading order in an expansion in powers of the small quark masses over the QCD scale, which we approximately identify with the proton mass $m_p$, we estimate $d_n/e=c_n[{(m_um_d)/(m_u+m_d)}]\bar\theta/m_p^2$ with $c_n$ expected to be of order unity. An explicit calculation in chiral perturbation theory suggests $|c_n|$ may be as large as $\sim10$.~\footnote{There is in fact an enhancement due to a chiral log and a numerical factor slightly larger than expected by naive dimensional analysis. The authors of \cite{Crewther:1979pi} find $|c_n|\sim8$. That result is not the full QCD prediction, however, since it does not include the contribution of the neutron-dipole counter-term, which is of course incalculable within chiral perturbation theory.} Currently, the most stringent bound on the neutron electric dipole moment reads $|d_n|/e<1.8\times10^{-26}$ cm at $90\%$ CL~\cite{Abel:2020gbr}. The corrections to $d_n$ from the CKM phase are so much smaller that $d_n$ can be interpreted as being dominated by $\bar\theta$. Allowing for $|c_n|$ to be in the range $1-10$, the experimental bound translates roughly into
\ba\label{thetaBound}
|\bar\theta|<{\cal O}(0.5-5)\times10^{-10}.
\ea
In the near future the experimental sensitivity is expected to improve by a factor of order ten.

Now, the SM is only the low energy manifestation of a more fundamental description of the subatomic world. The ultimate theory is expected to reproduce the SM parameters and its particle content without fine-tunings, very much like the SM itself explains the values of the pion mass and the muon lifetime. However, we have reviewed above that experiments clearly indicate that the pattern of CP violation in the SM is highly non-generic. This suggests that CP violation in the UV completion must also be non-generic. The UV completion of the SM must apparently feature sizable CP-odd phases, in order to explain $J$, and simultaneously justify \eqref{thetaBound}, namely the absence of CP-violation in the strong sector. {\emph{What properties should the UV completion of the SM have in order to accommodate these experimental facts without having to fine-tune its parameters?}} Identifying UV theories with the correct characteristics is so challenging that a specific term has been coined: the Strong CP Problem. Symmetry-based solutions of the Strong CP Problem include the QCD axion \cite{Peccei:1977hh,Weinberg:1977ma,Wilczek:1977pj}, the massless up-quark \cite{masslessU,Choi:1988sy,Banks:1994yg}, generalized parity \cite{Babu:1989rb,Barr:1991qx,Hook:2014cda}, and CP-invariant scenarios. In this paper we will focus on the latter.

Models belonging to the last class assume CP is exact in the UV and spontaneously broken in some hidden sector. The non-trivial task is to communicate the breaking to the SM so as to guarantee that \eqref{Jmeasured} and \eqref{thetaBound} are satisfied. The first concrete step towards viable scenarios of this type was achieved in \cite{Nelson:1983zb} and \cite{Barr:1984qx}, where a class of models with a tree-level CKM phase and a loop-induced $\bar\theta$ were identified. Most of the earlier constructions however suffer from unacceptably large 2-loop effects to the topological angle \cite{Vecchi:2014hpa}. Here we will investigate in detail the predictions of the few scenarios that can structurally ensure \eqref{thetaBound}, which we call models of $d-$mediation and $u-$mediation. Other recent work on scenarios of spontaneous CP breaking include \cite{Dine:2015jga,Albaid:2015axa,Davidi:2017gir,Ohmer:2018ali,Evans:2020vil,Schwichtenberg:2018aqc,Choi:2019omm,Perez:2020dbw,Cherchiglia:2020kut,Cherchiglia:2021vhe}. 

We begin presenting the effective field theory description of the minimal models of $d-$ mediation and $u-$mediation in Section \ref{sec:key}. There we also point out a hidden and subtle coincidence of scales that these scenarios must feature (see Section \ref{sec:puzzle}). UV completions of these models can explain such coincidence, as will be shown in a companion paper \cite{ValentiVecchi}. In this work this coincidence will be taken for granted and the focus will be on the phenomenology. A crucial step in assessing the viability of the models is estimating the irreducible corrections to $\bar\theta$, namely those arising from the defining ingredients of these scenarios. This is done in Section \ref{sec:irreducible}, where both non-decoupling and decoupling contributions to $\bar\theta$ are analyzed. The former dominate as the new physics mass is increased, whereas the latter may be relevant for masses close to the weak scale. A generalization to scenarios with several mediators is presented in Section \ref{sec:family}. Constraints from direct searches, electroweak tests and flavor violation are collected in Section \ref{sec:collider}. Our conclusions are presented in Section \ref{sec:conclusion}. The appendix contains a few technical details necessary to carry out our analysis.

\section{$d$- and $u$-Mediation}
\label{sec:key}

The basic assumption underlying the approach to the Strong CP Problem we study in this paper is that CP is a good symmetry of the UV. This means that in the effective field theory below some UV cutoff there exists a field basis in which all topological angles vanish and all couplings are real. CP is spontaneously broken by a color-neutral sector, whose relevant degrees of freedom are CP-odd scalars $\Sigma$ with non-vanishing vacuum expectation value, and then communicated to the SM via a mediator sector of $\psi$ particles characterized by a (CP-conserving) mass scale $m_\psi$. To keep the radiative corrections to the QCD theta angle under control, the CP-odd scalars should interact with the SM only via couplings that carry spurionic charges under the SM flavor symmetry; these couplings cannot involve the doublet quark representation ($q$) \cite{Vecchi:2014hpa}. Within a perturbative framework, these requests are so significant that basically leave us with two (non-exclusive) options. Either $\Sigma$ couple to the singlet up ($u$) or the singlet down ($d$) quark representations. We will refer to these scenarios as models of $d-$ and $u-$mediation. It is also possible to consider linear combinations of these two.

Explicitly, the Lagrangian of the minimal models with $d$-mediation consists of the obvious kinetic terms, including the gauge interactions, a potential $V(\Sigma,H)$ for the scalars, plus the following Yukawa and mass terms~\footnote{Throughout the paper we use a 2-component Weyl notation for the fermions.}
\ba\label{d-Mess}
-{\cal L}^d_{\rm Yuk}&=&y_uqHu+y_dq{\widetilde H}d\\\no
&+&y\psi \Sigma d+m_\psi\psi\psi^c+{\rm hc}~~~~~~~~(d{\rm-mediation}),
\ea
where $y_{u},y_d,y,m_\psi$ are CP-even matrices in flavor space, the theta angles vanish by CP, and $\psi$ ($\psi^c$) has SM charges conjugate (equal) to $d$. Summation over the family and gauge indices is understood. A completely analogous Lagrangian can be written for minimal models with $u$-mediation. The only difference is that $d$ in \eqref{d-Mess} is replaced by the electroweak singlet up representation and $\psi$ ($\psi^c$) should have SM charges conjugate (equal) to $u$:
\ba\label{u-Mess}
-{\cal L}^u_{\rm Yuk}&=&y_uqHu+y_dq{\widetilde H}d\\\no
&+&y\psi \Sigma u+m_\psi\psi\psi^c+{\rm hc}~~~~~~~~(u{\rm-mediation}),
\ea
We emphasize that in either case $\Sigma$ should be a family of at least two scalars. If there was just one scalar, the CP-odd phase in its vacuum expectation value could be removed from \eqref{d-Mess} via a re-definition of $\psi$ and $\psi^c$, so no CKM phase would be induced. The fermions $\psi,\psi^c$ could come in a family of fields, as well, though this is not strictly necessary (we will discuss this possibility in Section \ref{sec:family}). The scenario in \eqref{d-Mess} is a particular incarnation of the Nelson-Barr class \cite{Nelson:1983zb,Barr:1984qx}, first proposed in \cite{Bento:1991ez}. 

As we will demonstrate in Section \ref{sec:puzzle}, scenarios of $d-$ and $u-$mediation ensure that the corrections to the QCD theta parameter arising from the messenger sector are sufficiently small. These are {\emph{irreducible corrections}}, since they are due to the very same messenger sector that is responsible for transferring CP violation to the SM, i.e. generating the CKM phase. To avoid larger contributions to $\bar\theta$, some symmetry must be invoked to forbid $q\widetilde H\psi^c$ in \eqref{d-Mess} (or $qH\psi^c$ in models with $u$-mediation). On the other hand, interactions of the CP-violating sector with the leptons are basically unconstrained phenomenologically. 

Other, {\emph{reducible corrections}} to $\bar\theta$ generically arise from loops involving additional states, but these can all be naturally suppressed \cite{Nelson:1983zb}. As a concrete example let us consider the corrections to $\bar\theta$ from loops involving excitations of the CP-violating sector in the scenarios \eqref{d-Mess}. All symmetries of the theory allow a quartic scalar interaction with the Higgs doublet $\lambda_{mn}\Sigma^\dagger_m\Sigma_n|H|^2\subset V(\Sigma,H)$. Once this is taken into account it is easy to verify that one obtains 1-loop corrections of the type~\cite{Bento:1991ez}
\ba\label{1-loopTheta}
\bar\theta=\frac{c_y}{16\pi^2}\,\,{\rm Im}\left[\langle\Sigma\rangle^\dagger\lambda\frac{1}{m_\Sigma^2}y^*y^t\langle\Sigma\rangle\right]
\ea
for some real number $|c_y|\sim1$. These would be unacceptably large for generic couplings of order one. However it is possible to naturally suppress \eqref{1-loopTheta} taking $|y|\ll1$. Completely analogous considerations hold for other {\emph{reducible}} contributions to $\bar\theta$, such as those from the gauge sector of the original model of \cite{Nelson:1983zb}. 

In the rest of the paper we will study the models in eqs. \eqref{d-Mess} and \eqref{u-Mess} in the simplifying limit $|y|\ll1$. In this limit the fluctuations of the CP-breaking sector can be ignored, see \eqref{1-loopTheta}, and the corrections to $\bar\theta$ arise entirely from loops of the SM and the mediator sector. The latter effects are {\emph{irreducible}} and therefore fundamentally characterize this approach to the Strong CP Problem.

\subsection{Reproducing the SM: a Coincidence of Scales}
\label{sec:puzzle}

Models of $d-$ and $u-$mediation are very efficient at taming the irreducible corrections to $\bar\theta$, as we will see in Section \ref{sec:irreducible}. However, any construction of the type proposed in \cite{Nelson:1983zb,Barr:1984qx} must satisfy a highly non-trivial condition in order to reproduce the SM at scales below the messengers mass. Specifically, the two scales $y\langle\Sigma\rangle, m_\psi$ should be comparable to each other~\cite{Vecchi:2014hpa}
\ba\label{coincidence}
{\rm Im}(y\langle\Sigma\rangle)\sim{\rm Re}(y\langle\Sigma\rangle)\sim m_\psi,
\ea
within a few orders of magnitude, depending on the model. We will see shortly that $|y\langle\Sigma\rangle|\gtrsim |m_\psi|$ is necessary to generate a sizable CKM phase, whereas the more subtle $|y\langle\Sigma\rangle|\lesssim |m_\psi|$ is required to reproduce the quark mass spectrum within a reliable perturbative description. To appreciate the origin of \eqref{coincidence} we focus on the model in eq. \eqref{d-Mess} with $|y|\ll1$, as anticipated above. 
The generalization to scenarios with a family of $\psi$'s is presented in Section \ref{sec:family}.

The Yukawa part of the Lagrangian \eqref{d-Mess}, under our assumption that the only remnant of the CP-violating sector is the vacuum expectation value of $\Sigma$, can be written as
\ba\label{Ldfrozen}
\left.{\cal L}^d_{\rm Yuk}\right|_{\rm frozen}&=&-y_uqHu-y_dq{\widetilde H}d -\xi^\dagger\psi d-m_\psi\psi\psi^c+{\rm hc}\\\no
&=&-y_uqHu-
\left(
\begin{matrix}
q & \psi
\end{matrix}\right)
\left(
\begin{matrix}
y_d{\widetilde H} & 0\\
\xi^\dagger & m_\psi
\end{matrix}\right)
\left(
\begin{matrix}
d\\
\psi^c
\end{matrix}\right)+{\rm hc},
\ea
where to save typing we introduced the column vector $\xi$:
\ba\label{xiGen}
\xi^*_i=y_{mi}\langle\Sigma_m\rangle
\ea
In this setup CP violation is due entirely to $\xi$. In other words, there exists a field basis in which $y_u,y_d,m_\psi$ are real, the theta angles vanish, and the only complex quantity is $\xi$. 

Because $\psi,\psi^c$ are colored, collider bounds force them to lie above the TeV scale, see Section \ref{sec:collider}. It then makes sense to diagonalize the mass matrix neglecting the electroweak scale in a first approximation. Performing the following $SU(4)$ transformation
\ba\label{unitaryRot}
 \left(
\begin{matrix}
d\\
\psi^c
\end{matrix}\right)\to
\left(\begin{matrix}
{1} - \frac{\xi \xi^\dagger}{|\xi|^2}\left(1-\frac{m_\psi}{M}\right) & \frac{\xi}{M} \\
	-\frac{\xi^\dagger}{M}  & \frac{m}{M}
\end{matrix}\right)
 \left(
\begin{matrix}
d\\
\psi^c
\end{matrix}\right)
\ea
the Yukawa sector becomes
\ba\label{LdNew}
\left.{\cal L}^d_{\rm Yuk}\right|_{\rm frozen}&\to&-Y_uqHu-Y_dq{\widetilde H}d-Yq{\widetilde H}\psi^c-M\psi\psi^c+{\rm hc}\\\no
&=&-Y_uqHu-
\left(
\begin{matrix}
q & \psi
\end{matrix}\right)
\left(
\begin{matrix}
Y_d{\widetilde H} & Y{\widetilde H}\\
0 & M
\end{matrix}\right)
\left(
\begin{matrix}
d\\
\psi^c
\end{matrix}\right)+{\rm hc}
\ea
with (matrix multiplication is understood)
\ba\label{MYdY}
\begin{cases}
M^2={\xi^\dagger\xi+m_\psi^2}\\
Y_u=y_u\\
Y_d = y_d \left[{1}- \frac{\xi \xi^\dagger}{|\xi|^2} \left(1-\frac{m_\psi}{M}\right)\right]\\
Y=y_d \frac{\xi}{M}=Y_d \frac{\xi}{m_\psi}.
\end{cases}
\ea
After our unitary rotation \eqref{unitaryRot} the heavy degrees of freedom have real masses $M$ and complex Yukawa couplings $Y$. At scales $\ll|M|$ we integrate them out and recover the SM including higher-dimensional operators suppressed by inverse powers of $M$, see Section \ref{sec:collider}. The couplings $Y_u$ and $Y_d$ are the SM Yukawas, up to small loop effects. At the renormalizable level, CP violation is encoded in the complex matrix $Y_d$, or more precisely in a tree-level order CKM phase, and a radiatively generated theta angle $\bar\theta$. They will be our focus next.

\subsubsection{The CKM Phase and Perturbativity}
\label{sec:CKM}

A CKM phase compatible with \eqref{Jmeasured} can be reproduced as long as $\xi/m_\psi$ has imaginary and real entries of comparable order of magnitude satisfying
\ba\label{CKMcond}
\left|\frac{\xi}{m_\psi}\right|=|Y^{-1}_dY|\gtrsim1.
\ea
Before proving \eqref{CKMcond} let us attempt to derive the CKM phase in the limit $|\xi|\ll |M|$, where analytical calculations can be easily performed.

We go in the basis where $Y_u$ is diagonal, where the CKM matrix is the unitary matrix that diagonalizes $Y_dY_d^\dagger=y_d(1-\xi\xi^\dagger/M^2)y_d^\dagger$. For $\xi=0$ the latter is just the orthogonal matrix that diagonalizes $y_dy_d^\dagger$. Equivalently, the SM masses and mixing angles are determined by $Y_u=y_u$ and $Y_d= y_d$ and the CKM phase vanishes. We can thus express the corresponding real CKM matrix using the Wolfenstein parametrization with $\eta$ set to zero. For non-zero $|\xi\xi^\dagger|\ll M^2$ the mixing angles are corrected at order $|\xi|^2/M^2$ whereas a small $\eta\propto{\rm Im}[\xi\xi^\dagger/M^2]$ is generated. Since the invariant $J$ is linear in $\eta$, the real part of $\xi$ can only enter at subleading $|\xi|^4/M^4$ order. Now, plugging \eqref{MYdY} in \eqref{Jarlskog} and Taylor expanding $J$ in powers of $\xi\xi^\dagger/M^2$ we obtain, at leading order,
\ba\label{ourJ}
J
&=&A(1-\rho)\frac{m_s}{m_b}\lambda_C^4~{\rm Im}\left[\frac{\xi_2{\xi}^\dagger_3}{M^2}\right]\;\left[1+{\cal O}\left(\frac{|\xi|^2}{M^2}, \lambda_C\right)\right]\\\no
&=&4.1\times10^{-5}~{\rm Im}\left[\frac{\xi_2{\xi}^\dagger_3}{M^2}\right]\;\left[1+{\cal O}\left(\frac{|\xi|^2}{M^2},\lambda_C\right)\right].
\ea
In deriving \eqref{ourJ} we also exploited the numerical relations $m_d\sim m_b\lambda_C^4$, $m_s\sim m_b\lambda_C^2$, $m_u/m_t\sim \lambda_C^7$, $m_c/m_t\sim\lambda_C^4$ (the masses are renormalized at $\sim1$ TeV), and expanded in the Cabibbo angle. The factor of $m_s/m_b$ in \eqref{ourJ} originates from the fact that the phases in $Y_dY_d^\dagger$ are controlled by the off-diagonal elements of the hermitian matrix $y_d{\rm Im}[\xi\xi^\dagger/M^2]y_d^\dagger$, and so they disappear when $m_s/m_b,m_d/m_b\to0$.

Eq. \eqref{ourJ} suggests that eq. \eqref{Jmeasured} can be reproduced provided ${\rm Im}[{\xi_2\xi^\dagger_3}/{M^2}]\sim0.73$. But there is a serious problem with the estimate \eqref{ourJ}: it is not possible to satisfy $|\xi_2||\xi_3|/M^2\sim0.73$ compatibly with the constraint $|\xi|^2=M^2-m_\psi^2\leq M^2$! We should conclude that the value of $|\xi|/M$ needed to apparently reproduce the observed CKM phase with \eqref{ourJ} is too large for the perturbative expansion used to be reliable \cite{Hiller:2002um}. Some non-perturbative technique must be employed to determine whether these models can or cannot generate the CKM phase. And this is complicated by the fact that with $|\xi|\sim M$ the CKM mixing angles are not just functions of $y_d$, but can be significantly affected by $\xi/M$.

Fortunately there is a way to basically ``integrate out" $y_d$ from the problem and obtain necessary and sufficient conditions on $\xi$ for Nelson-Barr models to reproduce the CKM phase. The argument goes as follows. We want to explicitly compute \eqref{Jarlskog} using the tree-level approximation $Y_{u,d}^{\rm SM}=Y_{u,d}$. Employing eq. \eqref{MYdY} and simple algebraic manipulations we get 
\ba\label{detC1}
{\rm det}\left[H_u,H_d\right]
&=&{\rm det}\left[h_u,h_d-YY^\dagger\right]\\\no
&=&Y^\dagger\left[h_u,\left[h_u,h_d\right]^2\right]Y-Y^\dagger Y~Y^\dagger h_u\left[h_u,h_d\right]h_uY
\\\no
&-&Y^\dagger h_u^2Y~Y^\dagger\left[h_u,h_d\right]Y+Y^\dagger h_uY~Y^\dagger\left\{h_u,\left[h_u,h_d\right]\right\}Y,
\ea
where we defined $H_{u,d}=Y_{u,d}Y_{u,d}^\dagger$ and $h_{u,d}=y_{u,d}y_{u,d}^\dagger$. In the first line of \eqref{detC1}, $H_d=h_d-YY^\dagger$ follows from \eqref{MYdY}. The second equality is a consequence of the fact that, for any traceless matrix $C$, ${\rm det} C={\rm Tr}[C^3]/3$. In our case $C=\left[h_u,h_d\right]-\left[h_u,YY^\dagger\right]$, with ${\rm det}\left[h_u,h_d\right]=0$ because $h_{u,d}$ are CP even. The non-vanishing terms in ${\rm det} C$ are traces containing powers of $\left[h_u,YY^\dagger\right]$, and can therefore be written as $Y^\dagger fY$ for some anti-symmetric function $f$ of $h_{u,d}$. The next step towards our  necessary and sufficient conditions relies on the observation that the functions $f$ can be equivalently re-written in terms of $H_{u,d}$ by re-using $H_d=h_d-YY^\dagger$. Note that this replacement should be carried out uniquely in $f$, and not in the $Y,Y^\dagger$ of \eqref{detC1}, otherwise we would obviously get back to the left-hand side of \eqref{detC1}. This replacement leads us to an important relation, which we write as follows
\ba\label{detCconst}
{\rm det}\left[H_u,H_d\right]=I_{2,1}+Y^\dagger Y~I_{1,2}+Y^\dagger H_u^2 Y~I_{1,0}-Y^\dagger H_u Y~I_{1,1}~~~~~~~(d{\rm-mediation})
\ea
where
\ba\label{invInm}
I_{2,1}&=&Y^\dagger\left[H_u,\left[H_u,H_d\right]^2\right]Y
\\\no
I_{1,2}&=&Y^\dagger H_u\left[H_u,H_d\right]H_u Y
\\\no
I_{1,1}&=&Y^\dagger \left\{H_u,\left[H_u,H_d\right]\right\} Y
\\\no
I_{1,0}&=&Y^\dagger \left[H_u,H_d\right] Y.
\ea
Because, within a tree-level approximation, $Y_{u,d}$ are the SM Yukawa matrices, we can express them in a convenient form, say taking a diagonal $Y_u=\widehat Y_u$ and writing $Y_d=V^*\widehat Y_d$, where $V$ is the CKM matrix in the Wolfenstein parametrization.~{\footnote{The $^*$ is a consequence of our non-standard definition of the Yukawa interaction.} Then eq. \eqref{detCconst} must be interpreted as a {\emph{constraint}} on the coupling $Y$, or equivalently on $\xi/m_\psi$. As promised, the dependence on $y_d$ is included but implicit, i.e. $y_d$ has been integrated out.

One can then easily solve \eqref{detCconst} via numerical integration. The parameter $\xi$ is defined by 2 angles, a modulus and three phases. However the overall phase can be removed by a vector-like rotation of the mediators, so in practice only two of its phases are physical. Scanning over 300 angles and phases we obtain the $|\xi|/m_\psi$ Vs $\eta$ plot shown in the upper part of Fig. \ref{Jnumeric}. For models of $d-$mediation (where \eqref{detCconst} has been derived) we find that $|\xi|/m_\psi\gtrsim2$ is necessary. The analytical approximation \eqref{ourJ} works very well for $|\xi|\ll M$ but becomes inadequate rather quickly as $\xi$ increases and fails to account for the large spread seen in the upper part of Fig. \ref{Jnumeric}. The numerical analysis also shows that $\xi_1$ is not very important and that $\xi_2,\xi_3$ should be comparable in size and have large phases. The irrelevance of $\xi_1$ is an expected consequence of our choice of basis, since the last equation in \eqref{MYdY} says that $\xi_1$ appears in \eqref{detCconst} multiplied by the smallest of the down-type Yukawas or larger powers of the Cabibbo angle. The main players are clearly $\xi_2,\xi_3$, as anticipated by \eqref{ourJ}. To obtain the observed $J$ from \eqref{detCconst}, the CP contributions due to terms $\propto{\rm Im}[\xi\xi^\dagger/m_\psi^2]$ in the structures $Y^*_2f_{23}Y_3$ should win over the terms $\propto\eta$ in terms like $Y_3^*\lambda_C^2f_{23}Y_3$ (see, e.g., the expression of the invariant $I_{2,1}$ in Appendix \ref{sec:InvNum}). This requires
\ba\label{relY}
\frac{{\rm Im}\left[\xi_2\xi_3^*\right]}{\xi_3\xi_3^*}\frac{(\widehat{Y_d})_2}{(\widehat{Y_d})_3}\gtrsim\lambda_C^2.
\ea
Because we approximately have $(\widehat{Y_d})_2/(\widehat{Y_d})_3\sim\lambda_C^2$, we see that \eqref{relY} is satisfied for $|\xi_2|\sim|\xi_3|$. This condition is visible in the lower-left plot of Fig. \ref{Jnumeric}. Barring accidental correlations between $y_u,y_d$ and $\xi$, these findings imply that in a generic basis all entries in $\xi$ should be comparable and satisfy $|{\rm Re}[\xi]|\sim|{\rm Im}[\xi]|\gtrsim m_\psi$. This proves our claim \eqref{CKMcond}.

\begin{figure}[t]
\begin{center}
\includegraphics[width=.47\textwidth]{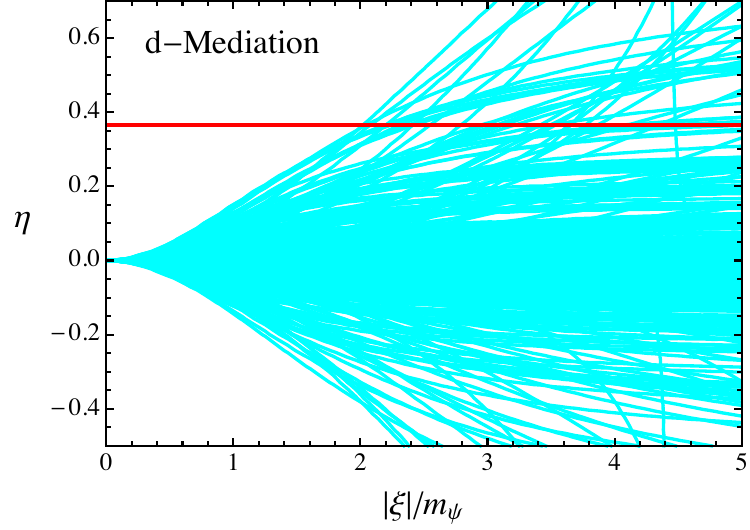}~~~~~~\includegraphics[width=.48\textwidth]{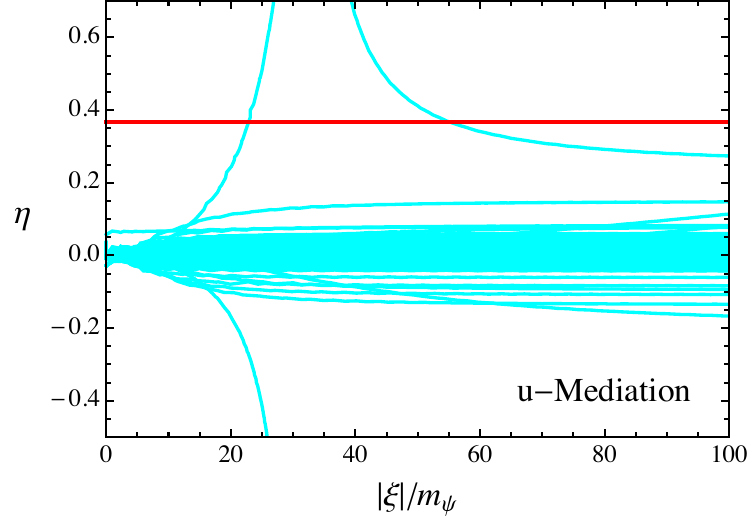}

\vspace{0.2cm}

\includegraphics[width=.47\textwidth]{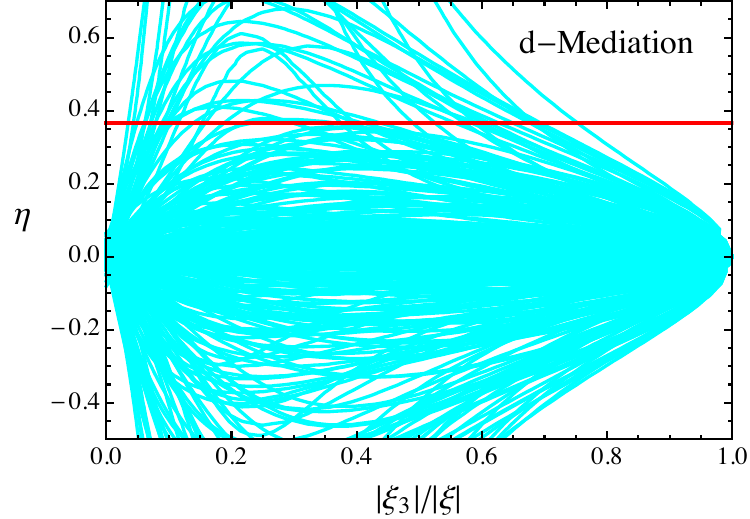}~~~~~~\includegraphics[width=.48\textwidth]{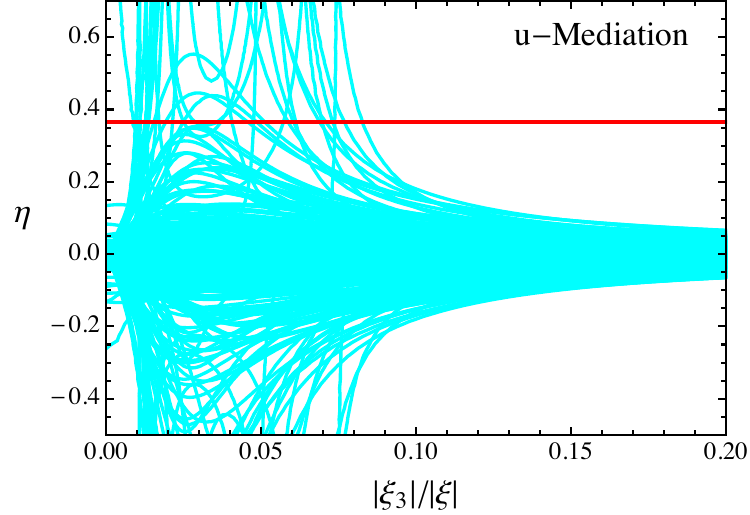}
\caption{\small Value of the CP-odd parameter $\eta$ of the Wolfenstein parametrization of the CKM (recall that $J=A^2\eta\lambda_C^6[1+{\cal O}(\lambda_C^2)]$) in models of $d-$mediation (left) and $u-$mediation (right). In the upper plots we generated 300 models with random values for the direction (2 angles) and the 2 physical phases of $\xi$, and kept an arbitrary dependence on the modulus $|\xi|/m_\psi$. Note the difference in the scale of the x-axis between the two plots. In the lower plots we also scanned over the modulus in the range $0<|\xi|/m_\psi<5$ ($d-$mediation), $20<|\xi|/m_\psi<40$ ($u-$mediation), and kept $|\xi_3|/|\xi|$ arbitrary. The red line indicates the real world value $J\simeq3.0\times10^{-5}$ and the cyan lines are the models' predictions.
\label{Jnumeric}}
\end{center}
\end{figure}

In $u-$mediation, a completely analog procedure leads to a relation similar to \eqref{detCconst} with the replacement $u\leftrightarrow d$. The field basis we adopt is now the one with $Y_d$ diagonal and $Y_u=V^t\widehat Y_u$. Only for very few choices of angles and $|\xi|/m_\psi\gtrsim20$ we can reproduce the CKM phase, as visible from the top-right plot of Fig. \ref{Jnumeric}. The basic reason can be traced back to the larger mass hierarchy of the up-quark sector. As a consequence, for example, in the $u-$mediation version of \eqref{ourJ} one finds a more significant $m_c/m_t$ suppression replaces $m_s/m_b$. As before, $\xi_1$ is irrelevant, but to satisfy the analog of \eqref{relY} one must have $|\xi_3|/|\xi_2|\lesssim m_c/(m_t\lambda_C^2)\sim0.07\sim\lambda_C^2$. This expectation is confirmed by the bottom-right plot of Fig. \ref{Jnumeric}. We conclude that models of $u-$mediation can reproduce the observed CKM phase provided their UV completion features some sort of anti-correlation between $y_u$ and $\xi$.

Eq. \eqref{CKMcond} is our first step towards \eqref{coincidence}. Analyzing the effective theory below $M$ carefully, one finds there is also an {\emph{upper bound}} on ${|\xi|}/{|m_\psi|}$. This is in fact a necessary condition if the effective theory has to reproduce the observed SM particle spectrum. When $|m_\psi|/|\xi|\to0$ the heavy state becomes a combination of $\psi$ and one component of the $d$'s, with a large Dirac mass $M\sim|\xi|$. The two orthogonal $d$ components have independent Yukawa couplings with only two $q$'s, whereas the remaining $q$  together with $\psi^c$ form a massless Dirac with an anomalous axial symmetry. This is the SM with a massless down-quark! We can confirm this observing that 
\ba\label{detYd}
{\rm det}(Y_d)={\rm det}(y_d)\frac{m_\psi}{M}, 
\ea
or perhaps more easily noting that in the limit $|m_\psi|/|\xi|\to0$ the SM Yukawa matrix $Y_d$ in \eqref{MYdY} becomes rank 2, $\lim_{|m_\psi|/|\xi|\to0}Y_d =  y_d \left[{1}- {\xi \xi^\dagger}/{|\xi|^2}\right]$. We see that the limit $m_\psi/|\xi|\to0$ is phenomenologically unacceptable. Clearly there should be a lower bound on $m_\psi/|\xi|$. Let us see what this is. For small but non-vanishing $|m_\psi|/|\xi|\ll1$ the mass of the down quark is of order $m_d\sim (\widehat{y}_dv/\sqrt{2})(m_\psi/|\xi|)$, where $\widehat y_d$ denotes a typical eigenvalue of $y_d$ and we used $M\sim|\xi|$. It is not possible to establish a firm bound on $m_\psi/|\xi|$ this way, however, because the value of $\widehat y_d$ is model-dependent and can in principle range between $(\widehat{Y}_d)_d$ and the non-perturbative $\sim4\pi$.


A robust, model-independent bound on $m_\psi/|\xi|$ can instead be derived from the UV description above $M$. Inspecting the field basis \eqref{LdNew} we see that the coupling $Y$ between the heavy fermionic state and the SM quark doublet becomes parametrically large when $|\xi|\gg |m_\psi|$, see the third line in \eqref{MYdY}. When $m_\psi$ is too small it becomes non-perturbative, say $|Y|>4\pi$, and we lose predictivity. Because Fig. \ref{Jnumeric} showed that $|\xi_i|\sim|\xi_j|$ is necessary to reproduce the CKM phase, the constraint $|Y|\ll4\pi$ may be expressed as 
\ba\label{non-pert}
\left|\frac{\xi}{m_\psi}\right|=|Y^{-1}_dY|\ll\frac{4\pi v}{\sqrt{2}m_b}\sim10^{3}~~~~~~~~~(d{\rm-mediation}).
\ea
Accidentally, this upper bound is numerically comparable to what one finds requiring the low energy theory reproduces the observed down-quark mass with $\widehat y_d\sim(\widehat{Y}_d)_b$, despite the two bounds have very different meaning. In the next subsection we will be able to quantify the perturbativity bound by inspecting the value of $\bar\theta$ predicted by these models.

In models with $u-$mediation ${|\xi|}/{|m_\psi|}\lesssim{m_t}/{m_u}\sim10^{5}$ is at least necessary to obtain a realistic spectrum if $\widehat y_u\sim(\widehat{Y}_u)_t$, from the effective field theory point of view. A stronger bound on $|\xi|/m_\psi$ applies however because we found that the condition $|\xi_3|\lesssim\lambda_C^2|\xi|$ is necessary to reproduce the CKM phase, see Fig. \ref{Jnumeric}. Hence, the UV description is non-perturbative unless $|Y|\ll4\pi$, or more explicitly $(\widehat Y_u)_c|\xi|/m_\psi\ll4\pi$ and simultaneously $\lambda_C^2(\widehat Y_u)_t|\xi|/m_\psi\ll4\pi$. The latter provides the most stringent bound, which reads
\ba\label{non-pert-u}
\left|\frac{\xi}{m_\psi}\right|=|Y^{-1}_uY|\ll\frac{4\pi v}{\sqrt{2}m_t\,\lambda_C^2}\sim300~~~~~~~~~(u{\rm-mediation}).
\ea
A concrete manifestation of this non-perturbativity problem is seen in potentially large radiative corrections to the $\bar\theta$ parameter when matching to the SM at scales $\sim M$, which we analyze below.

The coincidence \eqref{coincidence}, expressed more rigorously by the combination of Fig \ref{Jnumeric} and \eqref{non-pert} (or \eqref{non-pert-u}) cannot be explained within our effective field theory description. However it is important to appreciate that such relation is key to the viability of these models, since without it the low energy theory does not reduce to the SM. In the absence of a robust explanation of \eqref{coincidence} this solution of the Strong CP Problem is severely fine-tuned, and hence not convincing. It is reassuring that UV completions where \eqref{coincidence} naturally emerges exist, see \cite{ValentiVecchi}.

\subsection{Irreducible Contributions to $\bar\theta$}
\label{sec:irreducible}

Provided \eqref{CKMcond} and \eqref{non-pert} (or $|\xi|/m_\psi\gtrsim20$ and \eqref{non-pert-u} for $u-$mediation) are satisfied, the effective field theory below $M$ reproduces the SM up to irrelevant operators suppressed by inverse powers of $M$. The measured SM Yukawa couplings, including all radiative corrections, are given by
\ba\label{FuFd}
Y_u^{\rm SM}&=&F_u Y_u\\\no
Y_d^{\rm SM}&=&F_d Y_d,
\ea
where $F_{u,d}=1+{\cal O}(YY^\dagger, Y_{u,d}Y_{u,d}^\dagger)$ are 3 by 3 matrices functions of $Y_u, Y_d, Y$, $M$. The structure shown in \eqref{FuFd} may be understood taking advantage of the spurionic flavor charges of the SM Yukawas, for instance in the field basis in \eqref{LdNew}. That is, we interpret $Y,Y_u,Y_d$ as fields transforming under fictitious (spurious) flavor symmetries that leave \eqref{LdNew} invariant:
\ba\label{spurions}
Y_u&\to& U_q^*Y_uU_u^\dagger\\\no
Y_d&\to& U_q^*Y_dU_d^\dagger\\\no
Y&\to& U_q^*Y,
\ea
where $U_{q,u,d}$ are $SU(3)$ matrices. The Yukawas $Y_{u,d}^{\rm SM}$ in the effective field theory must be dimensionless combinations of the couplings of our theory that transform precisely as $Y_{u,d}$. This takes us to \eqref{FuFd}.

The SM topological angle, obtained by matching \eqref{LdNew} with the SM at the scale $M$, is given by \eqref{bartheta}:
\ba
\bar\theta&=&\theta-{\rm Im}~{\rm ln}({\rm det}[Y^{\rm SM}_u]{\rm det}[Y^{\rm SM}_d])\\\no
&=&\theta-{\rm Im}~{\rm ln}({\rm det}[F_u]{\rm det}[F_d]),
\ea
where we used that ${\rm det}[Y_{u,d}]$ are real, see \eqref{MYdY} and \eqref{detYd}. There are no tree-level contributions because the complex mass matrix of the colored fermions (see for example \eqref{Ldfrozen}) has real determinant. In other words, this model is in the Nelson-Barr class. Of course the same is true in \eqref{LdNew}, since the unitary matrix in \eqref{unitaryRot} has unit determinant. In our language, this just follows from the absence of tree-level flavor-invariant, CP-odd combinations of the parameters.

We can estimate the radiative contributions to $\bar\theta$ using the same trick as above. Contributions to $\bar\theta$ must be obviously CP-odd combination of our couplings but, importantly, also invariant under spurious $SU(3)$ rotations. This is because as we have seen its expression must be written in the flavor-invariant combinations $\theta, {\rm det}[F_u], {\rm det}[F_d]$. In addition, there can be a dependence on the electroweak scale $v\simeq246$ GeV, but for the moment let us work at leading order in the latter and set $v=0$.

The leading CP-odd flavor-invariant combination of our couplings is the one with fewer insertions of Yukawas. Up to a factor of order unity, and the appropriate power of the loop factor $1/16\pi^2$ needed to match the powers of couplings, such combination coincides with the value of $\bar\theta$ at the matching scale. With $Y=0$ we have the SM and the first correction to $\bar\theta$, analytic in the couplings, is very suppressed (see \cite{Khriplovich:1993pf} and the earlier literature \cite{Ellis:1978hq,Dugan:1984qf}). Potentially large corrections must involve $Y$. Eq. \eqref{spurions} shows that all the invariants are products of basic invariants built out of two powers of $Y$ and several $Y_u,Y_d$ (see Appendix~\ref{sec:InvNum} for a few examples). The CP-odd flavor-invariant with the smallest number of $Y_{u,d}$ insertions is (see $I_{1,0}$ in Appendix~\ref{sec:InvNum})
\ba\label{thetaDmodel}
\left.\bar\theta\right|_{\rm analy}&=&c_{\rm analy}
\left( \frac{1}{16\pi^2}\right)^3~{\rm Im}\left(Y^\dagger \left[ Y_d Y_d^\dagger, Y_u Y_u^\dagger \right] Y\right)\\\no
&\sim&\left( \frac{1}{16\pi^2}\right)^3~\lambda_C^2 {\widehat Y}_t ^2 {\widehat Y}_b ^3 {\widehat Y}_s \frac{{\rm Im}[\xi _i \xi _j ^*]}{m_\psi^2} \\\no
&\sim& 6 \times 10^{-18}\, \frac{{\rm Im}[\xi_i \xi_j^*]}{m_\psi^2}~~~~~~~~~(d{\rm-mediation}).
\ea
For the numerical estimate of \eqref{thetaDmodel} we went in the basis in which $Y_u$ is diagonal, where $Y_d$ is unitary up to CKM rotations, took $\lambda_C\sim0.23$ for the Cabibbo angle and renormalized the couplings at 1 TeV. The factor $\lambda_C^2$ arises because the result is proportional to the $23$ element of the CKM. In terms of familiar Feynman diagrams, this contribution to the QCD topological angle arises from 3-loop corrections to $F_{u,d}$ as well as direct corrections to $\theta$, with virtual fermions and the Higgs. Had we considered scenarios with unsuppressed couplings between the $q$'s and the messengers we would have found unacceptably large 2-loop corrections to $\bar\theta$ \cite{Vecchi:2014hpa}. 
Imposing $|\bar\theta|<10^{-10}$ on \eqref{thetaDmodel} one obtains a lower bound on $|m|/|\xi|$ a bit looser than \eqref{non-pert}. All other (subleading) flavor-invariants lead to weaker constraints. Non-perturbative values of the coupling $Y$ would imply unacceptably large corrections to $\bar\theta$.

Because in eq. \eqref{thetaDmodel} we neglected powers of $v$, that expression represents only the leading {\emph{non-decoupling}} contribution to $\bar\theta$, which dominates if $M$ is sufficiently large compared to the weak scale. When matching the UV theory to the SM at scales $\sim M$, however, one also finds additional threshold contributions that decouple as $16\pi^2v^2/M^2\to0$. To estimate the leading {\emph{decoupling}} effect we should allow $\bar\theta$ to depend on $v$, in which case its expression should respect the same selection rules \eqref{spurions} plus the additional spurious symmetry $v\to-v$, $(Y,Y_u,Y_d)\to-(Y,Y_u,Y_d)$. 

An important complication found when estimating the {{decoupling}} contributions is that these can be non-analytic in the couplings $Y_{u,d}$. Indeed, after electroweak symmetry breaking the Yukawas may appear not only as couplings, but also as masses. On the other hand, $\bar\theta$ is necessarily analytic in $Y$ because such a coupling does not control the large mass $M$ directly, but rather a small mixing angle of order $Yv/M$. We thus learn that the most general expression for $\bar\theta$, again leading in $Y$, reads
\ba
\bar\theta&=&{\rm Im}\left[Y^\dagger f\left(Y_d Y_d^\dagger, Y_u Y_u^\dagger,\frac{v^2}{M^2}\right) Y\right]\\\no
&=&{\rm Im}\left[Y^\dagger f_0\left(Y_d Y_d^\dagger, Y_u Y_u^\dagger,0\right) Y\right]+\frac{v^2}{M^2}{\rm Im}\left[Y^\dagger f_1\left(Y_d Y_d^\dagger, Y_u Y_u^\dagger,\frac{v^2}{M^2}\right) Y\right]+{\cal O}\left(\frac{v^4}{M^4}\right),
\ea
where $f$ is an unknown anti-symmetric 3 by 3 matrix, $f^t=-f$. The leading effect controlled by the term $f_0$ is precisely \eqref{thetaDmodel}. The main one proportional to $v^2/M^2$ is the decoupling effect we want to estimate. As explained above, $f_1$ can have a residual non-analytic dependence on $v^2$ that we cannot Taylor expand. Because this dependence is in principle arbitrarily complicated, it is not possible to find an explicit form of $f_1$ based solely on symmetry arguments. We can however reliably estimate the order of magnitude.

Anti-symmetry of $f$ requires that $f_1$ depends on both $Y_d Y_d^\dagger, Y_u Y_u^\dagger$. Some of this dependence could be hidden in logarithms of the masses; and these are precisely the quantities that are not constrained by our selection rules. Importantly, though, $f_1$ should be proportional to at least one power of $Y_d Y_d^\dagger$ and one power of $Y_u Y_u^\dagger$. The reason is that if the whole dependence of $f_1$ on, say, the up-type Yukawa was in the unknown non-analytic terms, then $\bar\theta$ would be singular in the limit $Y_u Y_u^\dagger\to0$. And this cannot be the case because such IR divergences do not appear in matching the Wilson coefficients of an effective theory. We conclude that the leading decoupling contributions scale similarly to \eqref{thetaDmodel} 
\ba\label{nonAnalyWZestSpurion}
\left.\bar\theta\right|_{\rm nonanaly}&\sim&c_{\rm nonanaly}\left(\frac{1}{16\pi^2}\right)^2~\frac{v^2}{M^2}\lambda_C^2 {\widehat Y}_t ^2 {\widehat Y}_b ^3 {\widehat Y}_s \frac{{\rm Im}[\xi _i \xi _j ^*]}{m_\psi^2}
\\\no
&\sim&c_{\rm nonanaly}~5\times10^{-17}\left(\frac{\rm TeV}{M}\right)^2\frac{{\rm Im}[\xi_i\xi_j^*]}{m_\psi^2}~~~~~~~~~(d{\rm-mediation}),
\ea
up to logarithms of the masses that we cannot estimate using spurion techniques, which have been included in $c_{\rm nonanaly}$. In Appendix \ref{sec:massbasis} we perform a more standard loop analysis in the mass basis and confirm this result, see \eqref{nonAnalyWZest}.

The powers of $1/16\pi^2$ in \eqref{nonAnalyWZestSpurion} are different from those of \eqref{thetaDmodel} because the decoupling effect is proportional to a mass squared (or analogously to $\propto v^2/M^2$), rather than a coupling squared. One can easily appreciate why the number of loops (and hence $1/16\pi^2$'s) in \eqref{nonAnalyWZestSpurion} is one less than \eqref{thetaDmodel} by re-instating the powers of $\hbar$ and observing that the Yukawa couplings scale as $\sim\hbar^{-1/2}$, whereas $v^2\sim\hbar$, $M\sim\hbar^0$. As a result, at least at the leading order, the main qualitative difference between non-decoupling \eqref{thetaDmodel} and decoupling \eqref{nonAnalyWZestSpurion} contributions is the formal replacement ${1}/({16\pi^2})\to {v^2}/{M^2}$, which indicates that decoupling effects are parametrically less relevant than the non-decoupling ones when $M\gtrsim4\pi v\sim3$ TeV. More precisely, the experimental bound \eqref{thetaBound} reads $|\xi|/m_\psi\lesssim(800-900)(M/{\rm TeV})$ for numbers $c_{\rm nonanaly}$ of order unity. 

Yet, we argued above that $c_{\rm nonanaly}$ may contain large logs of the mass ratios (dimensional analysis is enough to show this does not occur in $c_{\rm analy}$). Those that may in principle impact our estimate are the logs of the largest available mass, namely $M^2$. Inspecting the relevant diagrams one sees there are at most 3 powers of large logs. This is compatible with the SM computation of \cite{Khriplovich:1993pf}, which is one loop higher. As a conservative estimate, we may thus take $c_{\rm nonanaly}\sim\ln^3 M^2/m_b^2$ and the bound becomes $|\xi|/m_\psi\lesssim30(M/{\rm TeV})$. In Section \ref{sec:collider} we will see that electroweak constraints lead to bounds on the very same quantity that are comparable to this one, and obviously much more accurate theoretically than our order one estimate \eqref{nonAnalyWZestSpurion}. A genuine 2-loop computation would be necessary to determine the value of $c_{\rm nonanaly}$ as well as whether \eqref{nonAnalyWZest} can realistically compete with the bounds of Section \ref{sec:collider}. This calculation is however beyond the scope of the present paper.

In models with $u-$mediation, repeating an analysis completely analogous to the one leading to \eqref{thetaDmodel}, we find   
\ba\label{thetaUmodel}
\left.\bar\theta\right|_{\rm analy}&\sim&
\left( \frac{1}{16\pi^2}\right)^3~{\rm Im}\left(Y^\dagger \left[ Y_d Y_d^\dagger, Y_u Y_u^\dagger \right] Y\right)\\\no
&\sim&\frac{\lambda_C^2 {\widehat Y}_b ^2 {\widehat Y}_t ^3 {\widehat Y}_c}{(16\pi^2)^3} \frac{{\rm Im}[\xi _2 \xi _3 ^*]}{m_\psi^2} \\\no
&\sim& 5 \times 10^{-15}\, \frac{{\rm Im}[\xi_2 \xi_3^*]}{m_\psi^2}~~~~~~~~~(u{\rm-mediation}).
\ea
As it was for models of $d-$mediation, $|\bar\theta|<10^{-10}$ gives a constraint consistent with the perturbative bound, see \eqref{non-pert-u}, but a bit milder (recall that $|\xi_3|\sim\lambda_C^2|\xi|$ here). Analogously to \eqref{nonAnalyWZestSpurion}, decoupling corrections are of the same order as \eqref{thetaUmodel} up to the replacement ${1}/({16\pi^2})\to {v^2}/{M^2}$ and possibly large logs. The bound on $|\xi/m_\psi|$ from the electroweak $T$ parameter analyzed in Section \ref{sec:collider} is stronger and more accurate.

\subsection{Generalization to More Families of Mediators}
\label{sec:family}

The results of Sections \ref{sec:puzzle} and \ref{sec:irreducible} can be generalized to the case in which the mediators appear in different families with index $a,b=1,\cdots, n_\psi$. We limit our analysis to scenarios where $m_\psi$ has non-degenerate eigenvalues and all the mediators $\psi_a$, in the basis in which $m_\psi$ is diagonal, mix with the SM fermions. This is equivalent to saying that
\ba\label{xiGenGen}
\xi^*_{ia}=y_{mia}\langle\Sigma_m\rangle
\ea
is non-vanishing for any $i$ when $m_\psi$ is diagonal. The special rotation that removes the mass mixing (before electroweak symmetry breaking) now is
\ba\label{unitaryRotGen}
 \left(
\begin{matrix}
d\\
\psi^c
\end{matrix}\right)\to
\left(\begin{matrix}
A & {\xi}{(M^\dagger)}^{-1} \\
	-m_\psi^{-1}{\xi^\dagger}A  & m_\psi^\dagger {(M^\dagger)}^{-1}
\end{matrix}\right)
 \left(
\begin{matrix}
d\\
\psi^c
\end{matrix}\right)
\ea
where the condition $AA^\dagger=1-\xi(MM^\dagger)^{-1}\xi^\dagger=[1+\xi(m_\psi m_\psi^\dagger)^{-1}\xi^\dagger]^{-1}$ is necessary to ensure this transformation is unitary (the second equality is a consequence of the first and our definition of $M$, see below). After the above rotation is performed the Yukawa and mass terms look formally as in \eqref{LdNew}, where summation over indices is always understood. The masses and couplings of \eqref{LdNew}, in matrix notation, now explicitly read
\ba\label{MYdYGene}
\begin{cases}
MM^\dagger={\xi^\dagger\xi+m_\psi m_\psi^\dagger}\\
Y_u=y_u\\
Y_d = y_d A\\
Y=y_d \xi({M^\dagger})^{-1}=Y_d A^{-1}\xi({M^\dagger})^{-1}.
\end{cases}
\ea
In models with $u-$mediation the very same results hold except for the replacement $d\leftrightarrow u$. The reminder of this subsection applies to both scenarios.

An analysis similar to the one performed in Section \ref{sec:CKM} says that ${\rm Im}[(AA^\dagger)_{ij}]\gtrsim1$ is necessary to reproduce \eqref{Jmeasured}. This of course means that ${\rm Im}[\xi\xi^\dagger]\gtrsim |m_\psi|^2$. Also, $|m_\psi|\ll|\xi|$ would signal a non-perturbative regime. To see this we multiply $AA^\dagger$ on the left by $\xi^\dagger$ and on the right by $\xi$, and use the definition of $MM^\dagger$, to obtain 
\ba\label{xiAAxi}
\xi^\dagger AA^\dagger\xi=m_\psi m_\psi^\dagger[1-(MM^\dagger)^{-1}(m_\psi m_\psi^\dagger)].
\ea
From this follows that if one eigenvalue of $m_\psi m_\psi^\dagger$ is much smaller than $|\xi^\dagger\xi|$ the matrix $AA^\dagger$ develops a null vector or, in other words, the rank of $A$ becomes smaller than 3. To see this let us go in the basis in which $m_\psi$ is diagonal and suppose that $[m_\psi]_{\hat a}=0$ for some $a=\hat a$. Eq. \eqref{xiAAxi} then reads $[\xi^*]_{i\hat a}[AA^\dagger]_{ij}[\xi]_{j\hat a}=0$, which implies $[\xi]_{j\hat a}$ is a null eigenvector because of our hypothesis \eqref{xiGenGen}. We may rephrase this stating that ${\rm det}[m_\psi]=0$ implies ${\rm det}[A]=0$. In the same limit the last equation in \eqref{MYdYGene} shows that $Y$ becomes non-perturbative. These considerations demonstrate that the coincidence $\xi\sim m_\psi$ (see \eqref{coincidence}) must be realized even in the general case with more families of mediators. With $n_\psi>1$ it is more appropriate to express it as 
\ba\label{xi/mgen}
\begin{cases}
2\lesssim|Y_d^{-1}Y|\ll10^3~&~(d{\rm-mediation})\\
20\lesssim|Y_u^{-1}Y|\ll300~&~(u{\rm-mediation}).
\end{cases}
\ea
As in Section \ref{sec:CKM}, $|Y_d^{-1}Y|$ (or $|Y_u^{-1}Y|$) must be at least of order one in order for $J$ to be reproduced and simultaneously cannot be much larger than quoted otherwise the theory becomes non-perturbative. The perturbative bounds are analogous to \eqref{non-pert} and \eqref{non-pert-u}. However, as opposed to what happened with a single family, we will now show that \eqref{thetaBound} sets more stringent upper limits than \eqref{xi/mgen}.

In the case $n_\psi>1$ the derivation of $\bar\theta$ deserves some care because the new family index allows us to build more flavor-invariants involving the mass matrix $M$. It is not immediately obvious that large corrections to the theta angle can be avoided. For brevity we will analyze the non-decoupling effects only; the decoupling corrections can be estimated as in the previous section. We will prove that there are no (non-decoupling) 2-loop corrections to $\bar\theta$ and that (non-decoupling) 3-loop contributions are under control, exactly as it was in the one-family models of Section \ref{sec:irreducible}. The basic ingredients, along with their spurious transformations, are
\ba
Y_u&\to& U_q^*Y_uU_u^\dagger\\\no
Y_d&\to& U_q^*Y_dU_d^\dagger\\\no
Y&\to& U_q^*YU_{\psi^c}^\dagger\\\no
M&\to& U_\psi^*MU_{\psi^c}^\dagger,
\ea
where $U_{q,u,d}$ and $U_{\psi,\psi^c}$ are $SU(3)$ and $SU(n_\psi)$ flavor matrices, respectively. As before, we are interested in corrections proportional to $Y$. To warm up, it is straightforward to see that there is no correction to $\bar\theta$ at 1-loop order. Indeed, the unique combination ${\cal O}(Y^2)$ that is invariant under rotations of the SM fermions is $Y^\dagger Y$. Similarly, the mass can only enter via $M^\dagger M$, which is invariant under $\psi$ rotations. Now, the class of flavor-singlets one can build out of these two objects are just traces of $Y^\dagger Y$ and (dimensionless functions of) $M^\dagger M$. Anything of this form will be automatically real and CP-even, however, since such matrices are hermitian. Hence there cannot be 1-loop corrections to $\bar\theta$.

Let us then move to the 2-loop order, considering first only combinations of $Y,M$. For this task it is convenient to make use of some group theory. Since the building blocks are hermitian $n_\psi\times n_\psi$ matrices they can be expanded in a basis of $SU(n_\psi)$ generators $T^A$. Explicitly, 
\ba
h^AT^A\equiv Y^\dagger Y-\frac{1}{n_\psi}{\rm tr}[Y^\dagger Y],
\ea
and similarly for $M^\dagger M$. The trace parts are real and contribute to $\bar\theta$ at subleading order. Here we are interested in the leading contribution to the topological angle, so they can be safely neglected. With this notation, our 2-loop effects must be of the form
\ba\label{2loopYY}
h^Ah^B~F^{AB}(M^\dagger M),
\ea
where the dimensionless function $F^{AB}$ is symmetric and traceless. The key observation is that CP violation in the $SU(n_\psi)$ space acts as $T^A\to(T^A)^*=\eta^{AB}T^B$ on the generators and thus as $h^A\to \eta^{AB}h^B$ on the adjoints, where $\eta^{AB}$ can be chosen to be diagonal and satisfying $\eta^{AC}\eta^{CB}=\delta^{AB}$. Its explicit form depends on $n_\psi$. For example $\eta^{AB}={\rm diag}(1,-1,1)$ in $SU(2)$. From the algebra follows that the completely symmetric tensor $d^{ABC}$ is CP-even, whereas the structure function $f^{ABC}$ is CP-odd. We conclude that all $SU(n_\psi)$-invariant combinations of adjoints are automatically CP-even unless the expression contains an odd number of $f^{ABC}$. In particular, the combination \eqref{2loopYY} cannot contain the structure functions and is therefore CP-even: there is no 2-loop effect at ${\cal O}(Y^4)$. The absence of (non-decoupling) 2-loop contributions involving both $Y$ and $Y_{u,d}$ is even easier to understand. This class of flavor-invariants consists of traces of $Y_{u,d}Y_{u,d}^\dagger$, $YF(M^\dagger M)Y^\dagger$, which are again real by hermiticity, and hence CP-even.

We have thus demonstrated that there are no non-decoupling 2-loop contributions to $\bar\theta$ in these models. The first effects arise at 3-loops. Those due to $Y,M$ are constrained by the $SU(n_\psi)$ arguments given above. To get a non-vanishing combination at most one index of the structure function can be contracted with $h^A$ or a function of the masses, as these are all necessarily symmetric. The unique option is 
\ba\label{YYY}
h^Ah^Bh^CF^{A'B'C'}f^{AA'A''}f^{BB'B''}f^{CC'C''}d^{A''B''C''}, 
\ea
with $F^{A'B'C'}$ a dimensionless function of $M^\dagger M$. {(Note that the SM invariant ${\rm Im}~{\rm det}[Y_uY_u^\dagger,Y_dY_d^\dagger]$ is precisely of this form.)} A parametric estimate gives 
 \ba\label{YYY}
\left.\bar\theta\right|_{n_\psi\geq3}\sim
\begin{cases}
\left( \frac{1}{16\pi^2}\right)^3 {\widehat Y}_b^4{\widehat Y}_s^2\left({|Y_d^{-1}Y|}\right)^6\sim 10^{-21}\, \left({|Y_d^{-1}Y|}\right)^6~~~~&~~~~~(d{\rm-mediation})\\
\left( \frac{1}{16\pi^2}\right)^3 {\widehat Y}_t^4{\widehat Y}_c^2\left({|Y_u^{-1}Y|}\right)^6\sim 10^{-12}\, \left({|Y_u^{-1}Y|}\right)^6~~~~&~~~~~(u{\rm-mediation}),
\end{cases}
\ea
where we assumed all numerical factors are of order unity apart from the usual powers of $1/4\pi$. The bound \eqref{thetaBound} translates into much more stringent constraints than quoted in \eqref{xi/mgen}, because of the large powers of $Y$ involved. Importantly, though, this 3-loop contribution does {\emph{not}} exist if $n_\psi\leq2$, for the very same reason the SM with less than 3 generations has no Jarlskog invariant: the totally symmetric tensor vanishes and \eqref{YYY} cannot be built in those cases. More model-independent contributions, which exist also for $n_\psi=2$, must involve the SM Yukawas. The larger ones are proportional to the up-type Yukawa. The key building blocks are $Y^\dagger Y=h^AT^A+{\rm trace}$, $Y^\dagger Y_{u}Y_{u}^\dagger Y=h_u^AT^A+{\rm trace}$, and $F(M^\dagger M)=F^AT^A+{\rm trace}$. There is a unique way the indices of the CP-odd function $f^{ABC}$ can be contracted: $f^{ABC}h^Ah_u^BF^C(M^\dagger M)$. Including an appropriate number of the loop $1/16\pi^2$ factors, the latter CP-odd invariant can equivalently be written as
\ba\label{psi2Y6}
\left.\bar\theta\right|_{n_\psi\geq2}&\sim&\left( \frac{1}{16\pi^2}\right)^3~{\rm Im}~{\rm tr}\left(\left[Y^\dagger Y_{u}Y_{u}^\dagger Y,Y^\dagger Y\right]F(M^\dagger M)\right)
\\\no&\sim&
\begin{cases}
\left( \frac{1}{16\pi^2}\right)^3\lambda_C^2 {\widehat Y}_t ^2 {\widehat Y}_b ^3 {\widehat Y}_s\left({|Y_d^{-1}Y|}\right)^4\sim6\times10^{-18}\left({|Y_d^{-1}Y|}\right)^4  & (d{\rm-mediation})\\
\left( \frac{1}{16\pi^2}\right)^3 {\widehat Y}_t ^4 {\widehat Y}_c ^2\left({|Y_u^{-1}Y|}\right)^4\sim10^{-12}\left({|Y_u^{-1}Y|}\right)^4  & (u{\rm-mediation}).
\end{cases}
\ea
The numerical bound on ${|Y_d^{-1}Y|}$ (and ${|Y_u^{-1}Y|}$) following from \eqref{thetaBound} is a bit stronger than in the case of a single family of $\psi,\psi^c$, see \eqref{thetaDmodel} and \eqref{thetaUmodel}, again because of the larger power of the new Yukawa.

\section{Experimental Signatures}
\label{sec:collider}

The mediators $\psi,\psi^c$ are constrained by direct and indirect collider searches (for models of $d-$mediation see for example \cite{Cherchiglia:2020kut,Cherchiglia:2021vhe}). In this section we study both $d-$ and $u-$mediation and provide a qualitative assessment of the most stringent bounds for scenarios with one and two families of mediators, including those arising from radiative effects controlled by $Y$. We will see the latter are actually very relevant phenomenologically.

As shown in \eqref{MYdY} for a single family and \eqref{MYdYGene} for more families, $Y$ can be written as the $d-$type (or $u-$type) Yukawa multiplied on the right by a flavor-violating matrix. The latter is simply $[Y_d^{-1}Y]_{i}=\xi_i/m_\psi$ (or $[Y_u^{-1}Y]_{i}=\xi_i/m_\psi$) if a single family is considered, or more generally $[Y_d^{-1}Y]_{ia}$ (or $[Y_u^{-1}Y]_{ia}$). In models of $d-$mediation we have argued in Section \ref{sec:CKM} that $Y_d^{-1}Y$ must have entries of comparable size in order to reproduce the CKM. Therefore, for simplicity we treat it as a single coupling 
\ba
[Y_d^{-1}Y]_{ia}=|Y_d^{-1}Y|~~~~~~i=1,2,3
\ea
when quantifying the numerical bounds below. In other words, we will assume the only hierarchies involved in our calculations are those due to the quark masses and powers of the Cabibbo angle, and instead ignore possible cancellations in the sum of different $[Y_d^{-1}Y]_{ia}$'s.

The results of Section \ref{sec:CKM} reveal that in scenarios of $u-$mediation the CKM is reproduced provided $[Y_u^{-1}Y]_{ia}$ has a $i=3$ component suppressed by $\sim\lambda_C^2$ compared to the others. Our analysis will therefore be performed assuming that 
\ba\label{hypoUp}
[Y_u^{-1}Y]_{ia}=|Y_u^{-1}Y|\times\begin{cases}
1& i=1,2\\
\lambda_C^2 & i=3
\end{cases}
\ea
where $|Y_u^{-1}Y|$ is the parameter we will constrain, similarly to $d-$mediation. A more rigorous study of the phenomenology would require a numerical scan, but this is beyond the scope of our qualitative analysis.

Finally, when estimating the bounds on scenarios with two families for simplicity we take 
\ba\label{degenM}
M_1=M_2=M.
\ea
We will come back to the implications of this simplifying assumption below.

\subsection{$d$-Mediation}
\label{sec:dmen}

In models of $d-$mediation the $\psi,\psi^c$ behave similarly to a heavy b-quark: they are pair-produced and decay promptly into quarks and vector bosons or the Higgs boson. Current direct searches imply $m_\psi\gtrsim1400~{\rm GeV}$ for a single family~\cite{Sirunyan:2020qvb}.

The relevant couplings of the $\psi,\psi^c$ to the SM can be read off directly from \eqref{LdNew} (see also Section \ref{sec:family}). With the flavor indices shown explicitly, including those of $\psi,\psi^c$ ($a,b$), this is ${\cal L}^d_{\rm Yuk}\supset-Y_{ia}q_i{\widetilde H}\psi_a^c-M_{ab}\psi_a\psi_b^c$. Integrating out the heavy fermion at tree-level, below the scale $M_a$ we find (after a field re-definition of the quark doublet) a correction to the SM Lagrangian:
\ba\label{EFTSM}
\delta{\cal L}^{({\rm tree})}_{\rm SM}&=&\frac{1}{v^2}\left[{\bar c}_{ik}(Y_d)_{kj} q_i{\widetilde H}d_j |H|^2+{\rm hc}\right]\\\no
&-&\frac{1}{2v^2}{\bar c}_{ji}
\left[q^\dagger_i\bar\sigma^\mu q_j~H^\dagger i\overleftrightarrow{D _{\mu}}H+q^\dagger_i\sigma^a\bar\sigma^\mu q_j~H^\dagger \sigma^ai\overleftrightarrow{D _{\mu}}H\right]\\\no
{\bar c}_{ij}&=&\frac{v^2}{2}\left(Y\frac{1}{M^\dagger M}Y^\dagger\right)_{ij}.
\ea
An equivalent description of the following effects can be given in terms of the (flavor-violating) couplings in the mass basis (see Appendix \ref{sec:massbasis1}).

The second line of \eqref{EFTSM} gives rise to (flavor-violating) corrections to the $Z^0$-couplings of the down left-handed quarks. In the mass basis $Y$ reduces to $\widehat Y_d$ multiplied by an anarchic matrix on the right. The constraint on the flavor-diagonal components of $\bar c$ is therefore dominated by that on the bottom. In the mass basis it reads 
\ba\label{LEP}
|{\bar c}_{33}|\lesssim0.008,
\ea
(see e.g. \cite{Contino:2013kra} and note that in our model ${\bar c}_{ii}<0$). Other precision electroweak measurements are very weakly affected because of the small couplings involved. 

All couplings in \eqref{EFTSM} contribute to $\Delta F=1$ and $\Delta F=2$ transitions. Because of the strongly hierarchical structure inherited from the SM Yukawas, however, flavor-violating observables are far less crucial than in most scenarios of physics beyond the SM. The most constrained $\Delta F=2$ operators involve the left-handed down sector: 
\ba\label{dim6EFTFV}\label{treeFV}
{\cal L}_{\rm EFT}\supset -C^{d}_{ij;kl}~(d_L)_i^\dagger\bar\sigma^\mu (d_L)_j~(d_L)_k^\dagger\bar\sigma_\mu (d_L)_l.
\ea
The coefficient is corrected at tree-level $\delta C^{d}_{ij;kl}={\bar c}_{ji}{\bar c}_{lk}/(2v^2)$, from the second line of \eqref{EFTSM} via a $Z^0$ exchange, and also receives a 1-loop contribution from the exchange of the electroweak gauge bosons, of order $\delta C^{d}\sim\bar c g^2\lambda_C^2/(16\pi^2v^2)$. The main bound on \eqref{dim6EFTFV} in our model is currently due to $B_s-{\overline{B}}_s$ mixing, and conservatively reads $|C^d_{32;32}|\lesssim6.7\times10^{-12}~{\rm GeV}^{-2}$ (in the mass basis and when renormalized at $M\sim1$ TeV), see e.g. \cite{Silvestrini:2018dos}. The resulting bound on $|Y^{-1}_dY|/M$ is weaker than \eqref{LEP}. Operators in the $\Delta F=2$ class are also induced by the first line of \eqref{EFTSM}; however their coefficients are down by larger factors of the SM Yukawas. More importantly, at 1-loop the effective field theory below $M$ features additional $\Delta F=2$ interactions of the form ${\cal L}_{\rm EFT}\supset -C_{ij;kl}~q_i^\dagger\bar\sigma^\mu q_j~q_k^\dagger\bar\sigma_\mu q_l$, with coefficients (again in the mass basis in which $M^\dagger M$ is diagonal)
\ba\label{1loopFV}
C^{({\rm 1-loop})}_{ij;kl}=\frac{1}{8}\frac{1}{(4\pi)^2}Y^*_{ia}Y_{ja}Y^*_{kb}Y_{lb}\frac{\ln M_b^2/M_a^2}{M_b^2-M_a^2}.
\ea
The dominant constraints on these come again from $B_s-{\overline{B}}_s$ mixing and directly compete with \eqref{dim6EFTFV}. Importantly, \eqref{1loopFV} has a different parametric dependence than the corrections $\delta C^{d}$ mentioned above, i.e. it is controlled by $|Y^{-1}_dY|/\sqrt{M}$ rather than $|Y^{-1}_dY|/{M}$, and starts to dominate for masses $M\gtrsim4\pi v\sim3$ TeV. It even becomes more important than \eqref{LEP} at around $M\gtrsim18$ TeV.

Let us next move to $\Delta F=1$ observables. Among the most relevant operators are
\ba\label{DF=1}
{\cal L}_{\rm EFT}\supset (C_9)_{ij}(\overline d_L)_i\gamma^\mu (d_L)_j~\overline\ell\gamma_\mu \ell+(C_{10})_{ij}(\overline d_L)_i\gamma^\mu (d_L)_j~\overline\ell\gamma_\mu\gamma^5 \ell,
\ea
with $\ell$ any of the charged leptons and $(C_9)_{ij}=+{\bar c}_{ji}(1-4\sin^2\theta_w)/({2v^2})$, $(C_{10})_{ij}=-{\bar c}_{ji}/({2v^2})$. These follow from the second line of \eqref{EFTSM}, integrating out the $Z^0$, and contribute to rare $K$, $B$ meson decays, most notably $B\to X_s\ell\bar\ell$, $B_s\to\ell\bar\ell$, and $\epsilon'/\epsilon$. Conservatively requiring the new physics contribution lies within one sigma from the SM prediction (see, e.g.,~\cite{Aebischer:2019mlg,Bauer:2009cf}) we obtain bounds that are somewhat comparable numerically and a bit stronger than those derived from $B_s-{\overline{B}}_s$ mixing. However, they have the very same parametric dependence on $|Y^{-1}_dY|/M$ as \eqref{LEP}, and all turn out to be weaker than the latter. Loop-induced $\Delta F=1$ observables, say $B\to X_s\gamma$, lead to subleading bounds.

\begin{figure}[t]
\begin{center}
\includegraphics[width=.47\textwidth]{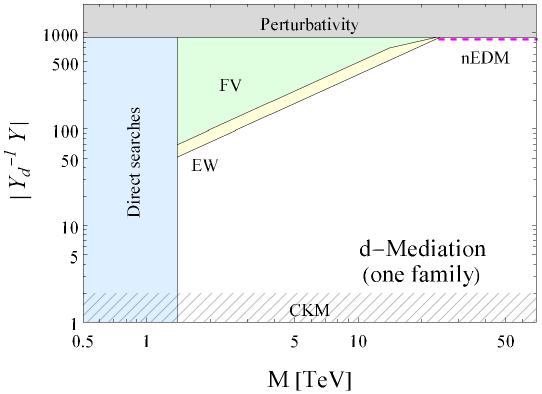}~~~~~~\includegraphics[width=.47\textwidth]{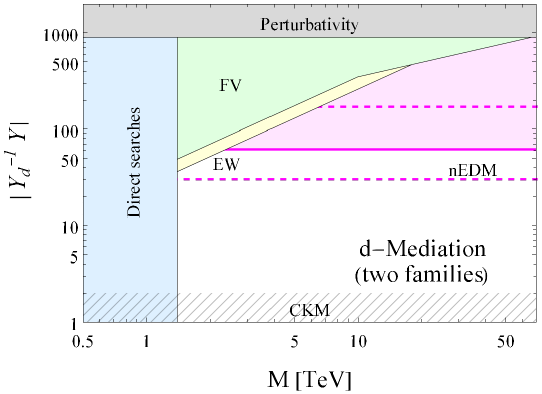}
\caption{\small Summary of the most relevant experimental constraints for models of $d-$mediation with one (left) or two (right) families of $\psi,\psi^c$. We show direct searches in light blue, electroweak bounds in yellow (``EW", see \eqref{LEP}), constraints from flavor violation in light green (``FV", in particular those due to the $\Delta F=1$ transitions discussed below \eqref{DF=1} are visible at lower mass and the $B_s-{\overline{B}}_s$ mixing constraint imposed on $C^{({\rm 1-loop})}$ dominates at larger $M$), and the bound from the neutron electric dipole moment in magenta (``nEDM") with a rough estimate of the error band (dashed lines). As seen in Section \ref{sec:CKM}, $|Y_d^{-1}Y|$ is also subject to a lower bound (see hatching), required in order for these models to be able to reproduce the CKM phase, and an upper bound, from perturbativity (see the grey region). See the text for details.\label{fig}}
\end{center}
\end{figure}

Finally we should not forget the constraints coming from the non-observation of the neutron electric dipole moment $d_n$. We estimated the dominant non-decoupling contributions to the $\bar\theta$ parameter for one family of mediators in \eqref{thetaDmodel} and for two families in \eqref{psi2Y6}. Because we did not compute the order one number in front of those expressions we allow for an unknown overall factor ranging within $[0.1,10]$. For one family only the most conservative bound, $|Y_d^{-1}Y|\lesssim860$, wins over the perturbative constraint \eqref{non-pert}. On the other hand, for two families of mediators \eqref{thetaBound} roughly translates into $|Y_d^{-1}Y|\leq60^{+100}_{-30}$, which is always stronger than \eqref{non-pert}. Other contributions to $d_n$ are induced by higher-dimensional operators. However, these give small corrections \cite{Vecchi:2014hpa}. For example, quark dipole interactions first arise at 2-loops and are suppressed by the small light quark masses.

As anticipated at the beginning of this section, we collect all the constraints in a single plot assuming that $Y_d^{-1}Y$ has anarchic entries of comparable size $|Y_d^{-1}Y|$. When plotting the bounds on the 2-family model we also take degenerate masses as in \eqref{degenM}. The mass degeneracy effectively increases the Wilson coefficients ${\bar c}_{ij}$ by a factor of $2$, thus making the electroweak and flavor constraints stronger than in the single family model.

The results are shown in Fig. \ref{fig} for one family of mediators (left) and two families (right). We see that direct searches as well as electroweak precision tests and flavor data have already started to constrain our scenarios, though a sizable portion of accessible parameter space around $M\gtrsim1$ TeV is still available. Remarkably, in the 2-family model the most significant constraint on the coupling constant comes from the 3-loop contribution to the neutron electric dipole moment. To make a more quantitative assessment of the allowed parameter space it would therefore be necessary to calculate the numerical coefficient in front of \eqref{psi2Y6}.

\subsection{$u$-Mediation}

In models of $u-$mediation collider searches currently imply $m_\psi\gtrsim1200~{\rm GeV}$ \cite{Aaboud:2018xuw}. In this regime it is appropriate to describe the electroweak and flavor constraints in terms of an effective field theory, as done for $d-$mediation.

The strongest constraint from electroweak observables arises due to radiative corrections to the $\widehat T$ parameter. In the effective field theory this corresponds to the dimensionless coefficient of $|H^\dagger D_\mu H|^2/v^2$. The main contributions are of order $Y^2\widehat Y_t^2$ and $Y^4$. Considering an arbitrary number of mediators' families, in the basis in which $M$ is diagonal we obtain: 
\ba\label{EWTP}
\widehat T&=&\frac{3}{16\pi^2} Y^*_{3a}Y_{3a}\,\frac{m_t^2}{M_a^2}\left(\ln \frac{M_a^2}{m_t^2}-1\right)\\\no
&+&\frac{3}{64\pi^2} \frac{v^2}{M^2}Y^*_{ia}Y_{ja}Y^*_{jb}Y_{ib}\,\frac{\ln M_b^2/M_a^2}{M_b^2-M_a^2}.
\ea
Since $S$ is very small we find that the $95\%$ CL bound of \cite{Baak:2012kk} simply reduces to $\Delta T=4\pi\widehat T/e^2<0.1$. Other electroweak observables lead to weaker bounds. In particular, corrections to the couplings of the up-type quarks to $Z^0$, most importantly those of the charm, are very small. 

The dominant $\Delta F=2$ effects show up in the radiative $K^0, B_d, B_s$ meson oscillations. The associated operator is again the one in \eqref{dim6EFTFV} with the Wilson coefficient \eqref{1loopFV}, where of course now $Y\propto Y_u$. Using respectively $|C^d_{21;21}|\lesssim2.0\times10^{-15}~{\rm GeV}^{-2}$, $|C^d_{32;32}|\lesssim6.7\times10^{-12}~{\rm GeV}^{-2}$, $|C^d_{31;31}|\lesssim8.0\times10^{-13}~{\rm GeV}^{-2}$ \cite{Silvestrini:2018dos} we find comparable bounds on $|Y_u^{-1}Y|/\sqrt{M}$, though $B_s-{\overline{B}}_s$ mixing slightly wins. This constraint is comparable to the one coming from \eqref{EWTP} at large $M$. Among the $\Delta F=1$ observables, by far the most constraining turns out to be $B\to X_s\gamma$. In the effective field theory the associated operator is first generated at 1-loop~\cite{Arnan:2016cpy}
\ba
{{\cal L}}_{\rm EFT}\supset-\frac{1}{9}\frac{e}{16\pi^2}\left(Y\frac{1}{M^\dagger M}Y^\dagger\right)_{23}~m_b~\overline{s_L}\sigma^{\mu\nu}b_R~F_{\mu\nu}.
\ea
The bound on $|Y_u^{-1}Y|/{M}$ derived from~\cite{Arnan:2016cpy} is weaker than the one due to the electroweak $T$ parameter as well as $B_s-{\overline{B}}_s$ mixing (already at $M\gtrsim100$ GeV).

\begin{figure}[t!]
\begin{center}
\includegraphics[width=.47\textwidth]{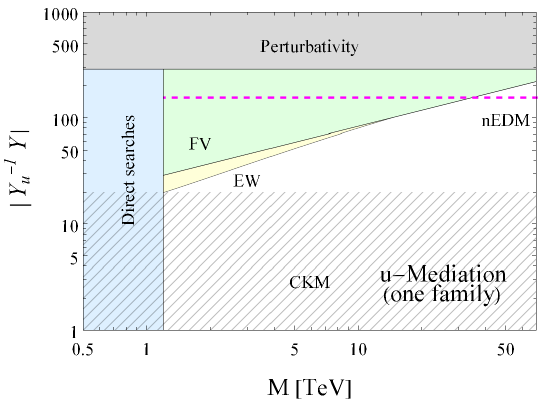}~~~~~~\includegraphics[width=.47\textwidth]{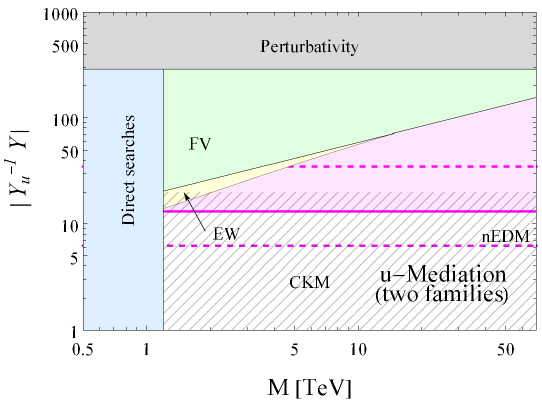}
\caption{\small Summary of the most relevant experimental constraints on models of $u-$mediation with one (left) or two (right) families of $\psi,\psi^c$. To reproduce the CKM we imposed the structure \eqref{hypoUp}. We show direct searches in light blue, electroweak in yellow (``EW", see \eqref{EWTP}), flavor violation in light green (``FV", from $B_s-{\overline{B}}_s$ mixing), the neutron electric dipole moment (``nEDM") in magenta. The lower and upper bounds from Section \ref{sec:CKM} are identified by the hatching and the grey region. See the text for details.\label{figUp}}
\end{center}
\end{figure}

Similarly to models of $d-$mediation, the constraints due to the neutron electric dipole are dominated by \eqref{thetaUmodel} for one generations of mediators and by \eqref{psi2Y6} for two generations. Following the same logic of Subsection \ref{sec:dmen} we find that in the former case only the most conservative bound on $\bar\theta$  is more stringent than the perturbativity requirement \eqref{non-pert-u}, whereas for two generations \eqref{thetaBound} translates, under our hypothesis \eqref{hypoUp}, into $|Y_u^{-1}Y|\leq13_{-7}^{+22}$.

All bounds are collected in Fig. \ref{figUp}. The main conclusions are similar to those drawn for theories of $d-$mediation. Yet, the parameter space of models of $u-$mediation is significantly reduced by the lower bound $|Y_u^{-1}Y|\gtrsim20$ necessary to reproduce the CKM phase, see Fig. \ref{Jnumeric}. In particular, $u-$mediation with two or more families appears to be basically excluded; a thorough numerical scan and an explicit computation of \eqref{psi2Y6} should tell us if some region of the parameter space is still allowed. Another, far less concrete, reason why $u-$mediation is less attractive is perhaps that $d-$mediation can be easily embedded into a grand-unified picture, where in terms of $SU(5)$ representations would be more appropriately called $\overline{\bf 5}$-mediation. On the other hand, $u-$type quarks come in the same multiplet as the doublets $q$, which we saw should not mix with the mediators otherwise $\bar\theta$ gets too large, and this generates a tension between models of ${\bf 10}$-mediation and \eqref{thetaBound}.

\section{Conclusions}
\label{sec:conclusion}

The Strong CP Problem remains one of the biggest challenges for physics beyond the Standard Model. In this paper we discussed an approach based on the idea of spontaneous CP violation. 

CP violation in the renormalizable SM is incapsulated into the CKM phase and the topological angles. There is a fundamental difference between these two. The CKM phase appears in flavor-violating observables whereas the topological angles control CP violation in {\emph{flavor-invariant}} processes. A robust way to produce a sizable CKM phase and a small QCD angle is therefore communicating the spontaneous violation of CP via flavor-violating couplings \cite{Vecchi:2014hpa}. Within a perturbative four-dimensional framework this can be achieved via complex mixings of the quarks with colored mediators. These are the so-called Nelson-Barr scenarios. Depending on the type of quark involved we speak of $q-$, $u-$ or $d-$mediation. The dominant source of CP violation however cannot come from models of $q-$mediation because in that case the radiative corrections to $\bar\theta$ are unacceptably large.

We analyzed Nelson-Barr models with $d-$ and $u-$mediation, where vector-like fermionic mediators mix respectively with the singlet $d-$ or $u-$quarks. We showed that these theories can reproduce the pattern of CP-violation as well as the fermion masses observed in Nature provided the two effective scales of the model, a CP-even mass $m_\psi$ and a CP-odd mass mixing $\xi$ between the mediators and the SM quarks, have comparable size. More quantitatively, combining the requirement \eqref{CKMcond}, necessary to have a sizable CKM (see Fig \ref{Jnumeric} for a more precise determination), with the perturbativity bounds \eqref{non-pert} or \eqref{non-pert-u} we have
\ba\label{coincidenceQ}
2\lesssim\left|\frac{\xi}{m_\psi}\right|\ll10^{3}~~~~~&&~~~~(d{\rm-mediation})\\\no
20\lesssim\left|\frac{\xi}{m_\psi}\right|\ll300~~~~~&&~~~~(u{\rm-mediation}).
\ea
It is important to stress that, while scenarios of $d-$mediation can generate the CKM phase with generic complex vectors $\xi$ with entries of comparable order, models of $u-$mediation require some sort of anti-correlation between $\xi$ and $y_u$. This makes such scenarios less generic than models of $d-$mediation. With several generations of mediators the coincidence \eqref{coincidenceQ} is more properly expressed as in \eqref{xi/mgen}.

Behind the perturbativity (upper) bound in \eqref{coincidenceQ} is the phenomenological requirement of reproducing the observed quark masses. Specifically, perturbative scenarios of $u-$mediation are precluded to have $m_\psi/|\xi|\ll1$ because in that regime they feature a very light up-quark, a mixture of one $q$ and $\psi^c$, of mass $m_u\sim ({\widehat y_uv}/\sqrt{2})(m_\psi/|\xi|)$. However we should note in passing that, if we allowed ourselves to speculate freely, we could address the Strong CP Problem by simply taking $m_\psi/|\xi|<10^{-10}~[m_u]_{\rm obs}\sqrt{2}/({\widehat y_uv})$ irrespective of whether CP was a good symmetry in the UV or not. That is, the very structure of the quark-mediator mixing characterizing Nelson-Barr scenarios realizes an interesting ``massless up-quark" solution of the Strong CP Problem, without flavor symmetries and with a sharp prediction: a heavy vector-like quark of mass $M\sim|\xi|$. All of this would be phenomenologically relevant if the massless up-quark solution was a viable option, which probably is not \cite{Aoki:2013ldr,Alexandrou:2020bkd}. In this paper we worked under the hypothesis that having a massless up-quark is not acceptable, and focused on scenarios that address the Strong CP Problem via spontaneous CP violation. It is following this reasoning that we derived the upper bound in \eqref{coincidenceQ}.

The coincidence \eqref{coincidenceQ} is a structural property of Nelson-Barr models, rooted in the way CP violation is communicated to the SM. Experimental constraints, including \eqref{thetaBound}, push the scales $|\xi|, m_\psi$ even closer. Such coincidence cannot be addressed via the effective field theory formalism employed here. Rather, it represents a constraint on the UV completion. In a subsequent paper we will show how \eqref{coincidenceQ} can emerge from realistic UV models \cite{ValentiVecchi}.

In scenarios with $d-$ and $u-$mediation, as opposed to other types of mediation, the irreducible contributions to the $\bar\theta$ parameter are suppressed up to an acceptable level. We showed that non-decoupling corrections to $\bar\theta$ do not arise before 3-loop order and decoupling contributions, non-analytic in the SM Yukawa couplings and potentially relevant for mediators' masses below $\sim4\pi v$, are generated already at 2-loop order. If a single family of mediators is considered, these effects are safely below the current bound \eqref{thetaBound} for parameters in the range \eqref{coincidenceQ}. The corrections to $\bar\theta$ become gradually larger as more families of mediators are added and result in more stringent constraints on $|\xi|/m_\psi$ than electroweak and flavor-violating observables as soon as the mediators are beyond the reach of direct searches. In the case of three or more families, for example, models of $d-$mediation are forced to have $|\xi/m_\psi|\lesssim50$, see \eqref{YYY}, whereas models of $u-$mediation in that case are basically excluded.

\section*{Acknowledgments}

We thank F. D'Eramo, M. K. Mandal, P. Mastrolia, P. Paradisi, M. Passera for discussions. This project has received support from the EU Horizon 2020 research and innovation programme under the Marie Sklodowska-Curie grant agreement No 860881-HIDDeN.

\appendix

\section{Flavor Invariants}
\label{sec:InvNum}

The invariants $I_{n,m}$ of \eqref{invInm} can be systematically expanded in powers of the Cabibbo angle employing the numerical relations $m_u/m_t\sim \lambda_C^7$, $m_c/m_t\sim\lambda_C^4$ and $m_d/m_b\sim \lambda_C^4$, $m_s/m_b\sim \lambda_C^2$. In the field basis in which the up Yukawa is diagonal ($Y_u=\widehat Y_u$ and $Y_d=V^*\widehat Y_d$) the leading terms are 
\ba
I_{2,1}&=&Y^\dagger\left[H_u,\left[H_u,H_d\right]^2\right]Y
\\\no&=& 2{\widehat Y}_t^4{\widehat Y}_c^2{\widehat Y}_b^4{\widehat Y}_s^2\left[1+{\cal O}\left(\lambda_C^2\right)\right]\\\no
&\times&\left[-A^2\eta\left(\frac{|\xi_2|^2}{m_\psi^2}+2\frac{|\xi_3|^2}{m_\psi^2}\right)\lambda_C^6
-A^2\eta\frac{{\widehat Y}_d}{{\widehat Y}_s}\lambda_C^5~{\rm Re}\left(\frac{\xi_2\xi_1^\dagger}{m_\psi^2}\right)
-A^2\eta\frac{{\widehat Y}_s}{{\widehat Y}_b}\lambda_C^4~{\rm Re}\left(\frac{\xi_3\xi_2^\dagger}{m_\psi^2}\right)
\right.
\\\no
&+&\left.A(1-\rho)\frac{{\widehat Y}_s}{{\widehat Y}_b}\lambda_C^4~{\rm Im}\left(\frac{\xi_3\xi_2^\dagger}{m_\psi^2}\right)
+A\frac{{\widehat Y}_d}{{\widehat Y}_b}\lambda_C^3~{\rm Im}\left(\frac{\xi_3\xi_1^\dagger}{m_\psi^2}\right)
+A^2\rho\frac{{\widehat Y}_d}{{\widehat Y}_s}\lambda_C^5~{\rm Im}\left(\frac{\xi_1\xi_2^\dagger}{m_\psi^2}\right)\right],
\ea
\ba
I_{1,2}&=&Y^\dagger H_u\left[H_u,H_d\right]H_u Y
\\\no&=& 2{\widehat Y}_t^4{\widehat Y}_c^2{\widehat Y}_b^4
\left[A\frac{{\widehat Y}_s}{{\widehat Y}_b}\lambda_C^2~{\rm Im}\left(\frac{\xi_3\xi_2^\dagger}{m_\psi^2}\right)\right]\left[1+{\cal O}\left(\lambda_C^2\right)\right],
\ea
\ba
I_{1,1}&=&Y^\dagger \left\{H_u,\left[H_u,H_d\right]\right\} Y
\\\no
&=& 2{\widehat Y}_t^4{\widehat Y}_b^4
\left[A\frac{{\widehat Y}_s}{{\widehat Y}_b}\lambda_C^2~{\rm Im}\left(\frac{\xi_3\xi_2^\dagger}{m_\psi^2}\right)\right]\left[1+{\cal O}\left(\lambda_C^2\right)\right],
\ea
\ba
I_{1,0}&=&Y^\dagger \left[H_u,H_d\right] Y
\\\no
&=& 2{\widehat Y}_t^2{\widehat Y}_b^4
\left[A\frac{{\widehat Y}_s}{{\widehat Y}_b}\lambda_C^2~{\rm Im}\left(\frac{\xi_3\xi_2^\dagger}{m_\psi^2}\right)\right]\left[1+{\cal O}\left(\lambda_C^2\right)\right].
\ea
In calculating \eqref{detCconst} the subleading terms are crucial because huge cancellations occur. For example, the invariants $I_{1,1}, I_{1,0}$ are not proportional to $m_c^2$ at leading order, but their sum in \eqref{detCconst} is. Similarly, several other important cancellations take place.

There are five more CP-odd flavor invariants one can build out of $H_{u},H_d$, and $Y$, but $I_{1,2}, I_{1,1}, I_{1,0}$ are the most relevant to our paper. Importantly, the largest ones in size (see $I_{1,0},I_{1,1}$) have a similar dependence on $A({{\widehat Y}_s}/{{\widehat Y}_b})\lambda_C^2~{\rm Im}({\xi_3\xi_2^\dagger}/{m_\psi^2})$. This explains the similarity in the factors in front of \eqref{thetaDmodel}, \eqref{nonAnalyWZestSpurion}, \eqref{psi2Y6}.

\section{In the Mass Basis}
\label{sec:massbasis}

\subsection{Diagonalization}
\label{sec:massbasis1}

After electroweak symmetry breaking there appear a new mass mixing of order $Yv/M$ in the down sector, see \eqref{LdNew}. We introduce the 4-family vectors
\ba\label{EWbasis}
D=
\left(
\begin{matrix}
q_d\\
\psi
\end{matrix}\right)~~~~~~~~~~
D^c=
\left(
\begin{matrix}
d\\
\psi^c
\end{matrix}\right)
\ea
and diagonalize the mass matrix via $SU(4)$ rotations $D\to U_{D}D$, $D^c\to U_{D^c}D^c$, 
\ba
U_D^t\left(
\begin{matrix}
\frac{Y_dv}{\sqrt{2}}  & \frac{Yv}{\sqrt{2}}\\
0 & M
\end{matrix}\right)U_{D^c}=\widehat M_{d}.
\ea
The resulting couplings to $W^\pm,Z^0$ and the Higgs boson $h$ (defined as $H^t=(0, v+h)/\sqrt2$ in the unitary gauge) read:
\ba\label{couplMassBasis}
{\cal L}_W&=&-\frac{g}{\sqrt{2}}V_{i\alpha}[q_u]_i^\dagger\bar\sigma^\mu D_\alpha~W^+_\mu+{\rm hc}
\\\no
{\cal L}_Z&=&-\frac{g}{2c_w}\left[[q_u]_i^\dagger\bar\sigma^\mu[q_u]_i -{\cal Z}_{\alpha\beta}D_\alpha^\dagger\bar\sigma^\mu D_\beta-2s_w^2J^\mu_{\rm em}\right]Z_\mu
\\\no
{\cal L}_h&=&-\frac{[\widehat M_u]_{ij}}{v}[q_u]_iu_j{h}-[{\cal Y}]_{\alpha\beta}D_\alpha D^c_\beta {h}+{\rm hc}
\ea
where $J^\mu_{\rm em}$ is the flavor-diagonal QED current of $q_u,u,D,D^c$, and 
\ba\label{VZM}
V_{i\alpha}&=&[U_{q_u}^\dagger]_{ij}[U_D]_{j\alpha}\\\no
{\cal Z}_{\alpha\beta}&=&[U_{D}^\dagger]_{\alpha i}[U_D]_{i\beta}\\\no
&=&\delta_{\alpha\beta}-[U_{D}^*]_{4\alpha}[U_D]_{4\beta}\\\no
[{\cal Y}]_{\alpha\beta}&=&\frac{[\widehat M_d]_{\alpha\beta}}{v}-\frac{M}{v}[U_D]_{4\alpha}[U_{D^c}]_{4\beta}.
\ea
Explicit expressions for $V,{\cal Z},{\cal Y}$ can be derived as an expansion in $Yv/M$ (see for example \cite{Cherchiglia:2020kut,Cherchiglia:2021vhe}). We show here only the leading order:
\ba\label{leadV}
U_D=
\left(
\begin{matrix}
U  & \frac{Y^*v}{\sqrt{2}M}\\
-\frac{Y^tv}{\sqrt{2}M}U & 1
\end{matrix}\right)\left[1+{\cal O}(Y^2v^2/M^2)\right],~~~~~U_{D^c}=
\left(
\begin{matrix}
U'  & 0\\
0 & 1
\end{matrix}\right)\left[1+{\cal O}(Y^2v^2/M^2)\right]
\ea
with $U^t(Y_dY_d^\dagger)U^*=U'^\dagger(Y_d^\dagger Y_d)U'=2\widehat M_d^2/v^2$.

\subsection{Decoupling Contributions to $\bar\theta$}

We describe here the irreducible, decoupling contributions to $\bar\theta$ in models of $d-$mediation. All estimates in this section are obtained working in the electroweak basis in which $Y_u$ is diagonal. In this case one effectively has $U_{q_u}=U_u=1$ and the expressions in \eqref{VZM} are all controlled by $V=U_D$ and $U_{D^c}$; also, at leading order the matrix $U$ should be identified with the 3-dimensional CKM matrix, see \eqref{leadV}.

First off, it is easy to see that there are no 1-loop corrections to $\bar\theta$, similarly to the SM. CP violation at 2-loops may arise due to loops of the $W^\pm$, $Z^0$ or the Higgs $h$. These can again contribute via corrections to the Yukawas \eqref{FuFd} or direct contributions to $\theta$. Inspecting the couplings of \eqref{couplMassBasis} we see that 2-loop diagrams with fermions and $W^\pm$ (we will refer to these as $W^\pm-W^\pm$ diagrams) generate corrections directly to $\theta$ of the form
\ba\label{nonAnalyWW}
\left.\bar\theta\right|_{{\rm nonanaly}, WW}&=&\left(\frac{g^2}{16\pi^2}\right)^2~{\rm Im}\left([V]_{i\alpha}[V^*]_{i\beta}[V]_{j\beta}[V^*]_{j\alpha}\right)~F^{ij;\alpha\beta}_{1WW}\\\no
&+&\left(\frac{g^2}{16\pi^2}\right)^2~{\rm Im}\left([V]_{i\alpha}[V^*]_{i\beta}[V]_{4\beta}[V^*]_{4\alpha}\right)~F^{i;\alpha\beta}_{2WW},
\ea
with $i,j=1,2,3$ and $\alpha,\beta=1,2,3,4$. Here $V_{i\alpha}$ a generalization of the CKM matrix and $F_{1WW,2WW}$ are real functions of the masses (squared) indicated by the corresponding indices. A sum over families is understood, though the imaginary prefactor is non-zero only when $i\neq j$ and $\alpha\neq \beta$. The functions $F_{1WW,2WW}$ are constrained by a number of physical considerations. By unitarity of the 4 by 4 matrix $V_{\alpha\beta}$, eq. \eqref{nonAnalyWW} vanishes unless $F_{1WW,2WW}$ depend on both $m^2_\alpha$ and $m^2_\beta$. Indeed, if the dependence on $m^2_\alpha$ was absent one could sum over $\alpha$ and obtain a vanishing expression because of a generalized GIM mechanism. Similar logic applies to $\beta$. On the other hand, if no dependence on $m^2_j$ (or equivalently $m^2_i$) exists we may replace $[V]_{j\beta}[V^*]_{j\alpha}=\delta_{\alpha\beta}-[V]_{4\beta}[V^*]_{4\alpha}$ in the above expression. The $\delta_{\alpha\beta}$ does not contribute but the reminder does not vanish. In practice, we can split eq. \eqref{nonAnalyWW} into a piece that has a non-trivial dependence on both $i\neq j$ ($F_{1WW}$) and one that does not depend on one of the two, say on $j$ ($F_{2WW}$).

Incidentally, the contribution $\propto F_{2WW}$ has exactly the same structure obtained in 2-loop corrections to $\bar\theta$ from diagrams with fermions, one $W^\pm$ and one $Z^0$ ($W^\pm-Z^0$ for short), up to an irrelevant correction:
\ba\label{nonAnalyWZ}
\left.\bar\theta\right|_{{\rm nonanaly}, WZ}=\left(\frac{g^2}{16\pi^2}\right)^2~{\rm Im}\left([V]_{i\alpha}[V^*]_{i\beta}[V]_{4\beta}[V^*]_{4\alpha}\right)~F^{i;\alpha\beta}_{WZ}.
\ea
To prove this observe that the $W^\pm-Z^0$ loop is proportional to (see \eqref{couplMassBasis})
\ba
&&\sum_\sigma{\rm Im}\left([V]_{i\beta}[V^*]_{i\gamma}[{\cal Z}]_{\sigma\gamma}[{\cal Z}^*]_{\sigma\beta}\right)~{F'}_{WZ}^{i;\beta\gamma\sigma}\\\no
&&={\rm Im}\left([V]_{i\beta}[V^*]_{i\gamma}[V]_{4\beta}[V^*]_{4\gamma}\right)\left[\sum_\sigma|V_{4\sigma}|^2{F'}_{WZ}^{i\beta\gamma\sigma}-{F'}_{WZ}^{i\beta\gamma\gamma}-{F'}_{WZ}^{i\beta\gamma\beta}\right]\\\no
&&\equiv{\rm Im}\left([V]_{i\beta}[V^*]_{i\gamma}[V]_{4\beta}[V^*]_{4\gamma}\right)F_{WZ}^{i;\beta\gamma}.
\ea
For the same reason explained above, $F_{WZ}$ must depend on $m^2_i,m^2_\beta,m^2_\gamma$ otherwise the sum vanishes. The resulting structure is the one in \eqref{nonAnalyWZ}, as promised. The loop diagrams with only fermions and $Z^0$ vanish because 
\ba
{\rm Im}\left([{\cal Z}]_{\alpha\beta}[{\cal Z}^*]_{\alpha\gamma}[{\cal Z}]_{\sigma\gamma}[{\cal Z}^*]_{\sigma\beta}\right)=0~~~~~~~~~~({\rm no~sum}),
\ea
as can be seen from the explicit expression of $[{\cal Z}]_{\alpha\beta}$ given in \eqref{couplMassBasis}.

Diagrams with virtual Higgs bosons also contribute, but we will see below they are subleading. Before showing this we discuss a key constraint that the functions $F_{1WW,2WW,WZ}$ are subject to. Because there cannot occur IR divergences in matching the UV to the SM effective field theory below the scale $M$, $F_{1WW,2WW,WZ}$ must be well-behaved when $M^2\gg m_{i,j}^2,m_{\alpha,\beta}^2, m_W^2$. Hence
\ba\label{nonAnalyMasses}
F_{1WW}^{ij;\alpha\beta}&=&{\rm min}\left(\frac{m_i^2}{m_{W}^2},\frac{m_j^2}{m_{W}^2}\right)\,{\rm min}\left(\frac{m_\alpha^2}{m_{W}^2},\frac{m_\beta^2}{m_{W}^2}\right)\,{\widetilde F}_{1WW}\left(\frac{m_i^2}{m_{\alpha}^2},\frac{m_j^2}{m_{\alpha}^2},\frac{m_\beta^2}{m_{\alpha}^2},\frac{m_W^2}{m_{\alpha}^2}\right)\\\no
F_{2WW}^{i;\alpha\beta}&=&\frac{m_i^2}{m_W^2}{\rm min}\left(\frac{m_\alpha^2}{m_{W}^2},\frac{m_\beta^2}{m_{W}^2}\right)~{\widetilde F}_{2WW}\left(\frac{m_i^2}{m_{\alpha}^2},\frac{m_\beta^2}{m_{\alpha}^2},\frac{m_{W}^2}{m_{\alpha}^2}\right)\\\no
F_{WZ}^{i;\alpha\beta}&=&\frac{m_i^2}{m_W^2}{\rm min}\left(\frac{m_\alpha^2}{m_{W}^2},\frac{m_\beta^2}{m_{W}^2}\right)~{\widetilde F}_{WZ}\left(\frac{m_i^2}{m_{\alpha}^2},\frac{m_\beta^2}{m_{\alpha}^2},\frac{m_W^2}{m_{\alpha}^2},\frac{m_{Z}^2}{m_{\alpha}^2}\right),
\ea
with the powers of $m_{i,\alpha}^2/m_W^2$ ensuring that $F_{1WW,2WW,WZ}$ be regular when {\emph{any}} of the quark masses involved vanishes. Of course this does not mean that the result is inevitably proportional to $m_u^2$ or $m_d^2$, because the CP-odd factor ${\rm Im}\left([V]_{i\alpha}[V^*]_{i\beta}[V]_{j\beta}[V^*]_{j\alpha}\right)$ is non-vanishing only for specific combinations of indices, the dominants of which do not necessarily involve the first generation. Actually \eqref{nonAnalyMasses} demonstrates that the dominant contributions to $\bar\theta$ arise from diagrams in which the heavier generations run in the loop. Finally, the end result for $\bar\theta$ should clearly be regular as $g\to0$. This tells us that ${\widetilde F}_{1WW,2WW,WZ}$ are analytic in the vector boson masses. The dominant contributions can be calculated for $m^2_W/M^2\to0$.

As a check of our arguments, one can verify that \eqref{nonAnalyWW} together with \eqref{nonAnalyMasses} reproduce the structure of the non-analytic contributions to $\bar\theta$ found in the SM \cite{Khriplovich:1985jr}. The unitarity of the CKM matrix however forces a non-trivial cancellation at 2-loops. Such degeneracy is lifted in diagrams with an additional strong coupling (or photon) loop, and so the actual end result is down by a factor $g_s^2/16\pi^2$ compared to \eqref{nonAnalyWW}. In our case no such cancellation takes place because there is no ``heavy top quark" to compensate for the fourth $d-$type family. For this reason $\bar\theta$ is already corrected at 2-loops.

We have all the tools to estimate the size of the non-analytic contributions to $\bar\theta$. We begin considering contributions to $F_{1WW}$, where both $i,j\neq i$ appear. The largest mass factor from \eqref{nonAnalyMasses} is obtained when $i=3, j=2$ and $\alpha=4,\beta=3$. The proportionality to $(m_c^2/m_W^2)(m_b^2/m_W^2)$ renders such corrections rather innocuous. In addition, \eqref{nonAnalyWW} turns out to be very small as well. Using the approximate expressions for $V$ derived in Appendix \ref{sec:massbasis}, and recalling that $Y$ can be written as a function of the SM Yukawa $Y_d$ and $\xi/m_\psi$ as in \eqref{MYdY}, we find that ${\rm Im}\left([V]_{34}[V^*]_{33}[V]_{23}[V^*]_{24}\right)\sim(m_b/M)\lambda_C^2(m_s/M)(|\xi|^2/m_\psi^2)$. Even including potentially large logs, the resulting contribution to $|\bar\theta|$ is at most numerically comparable to \eqref{thetaDmodel} for $M\sim1$ TeV. The effect becomes subleading with $({\rm TeV}/M)^2$ as soon as the mediator mass is above the TeV, which has to be the case because of direct searches (see Section \ref{sec:collider}).

Next we turn to an estimate of the $W^\pm-W^\pm$ loops controlled by $F_{2WW}$, or analogously of the $W^\pm-Z^0$ loop in eq. \eqref{nonAnalyWZ}, which we argued to be comparable parametrically. In this case the largest mass enhancement is obtained with $i=3$, $\alpha=4$, $\beta=3$, when the fermions running in the loop are the top, the bottom, and the heavy fermion. This is much larger than the effect proportional to $F_{1WW}$ that we just analyzed. On the other hand,  
\ba
{\rm Im}\left([V]_{34}[V^*]_{33}[V]_{43}[V^*]_{44}\right)
\sim\lambda_C^2\frac{m_bm_s}{M^2}\frac{{\rm Im}[\xi_2\xi_3^*]}{m_\psi^2}
\ea
is comparable to the imaginary part found above. The final result is thus expected to be of order
\ba\label{nonAnalyWZest}
\left.\bar\theta\right|_{{\rm nonanaly}, WZ}&=&c_{\rm nonanaly}\left(\frac{g^2}{16\pi^2}\right)^2~\frac{m_t^2}{m_{W}^2}\frac{m_b^2}{m_{W}^2}~\lambda_C^2\frac{m_bm_s}{M^2}\frac{{\rm Im}[\xi_i\xi_j^*]}{m_\psi^2}\\\no
&\sim&c_{\rm nonanaly}~10^{-16}\left(\frac{\rm TeV}{M}\right)^2\frac{{\rm Im}[\xi_i\xi_j^*]}{m_\psi^2}.
\ea
We discussed it below \eqref{nonAnalyWZestSpurion}. Here we just observe that an independent way to understand the necessity of factors of quark masses in front of \eqref{nonAnalyWZestSpurion} is to note that when all SM fermions are degenerate we can put $V_{ij}$ in diagonal form, in which case \eqref{nonAnalyWW} vanishes.

Loops with fermions and 2 virtual Higgses, or one Higgs and one $W^\pm$, or one Higgs and one $Z^0$ are respectively controlled by (no sum over indices is implied)
\ba
&&{\rm Im}\left([{\cal Y}]_{\alpha\beta}[{\cal Y}^*]_{\alpha\gamma}[{\cal Y}]_{\sigma\gamma}[{\cal Y}^*]_{\sigma\beta}\right)\\\no
&&{\rm Im}\left([V]_{i\beta}[V^*]_{i\gamma}[{\cal Y}]_{\gamma\sigma}[{\cal Y}^*]_{\beta\sigma}\right)\\\no
&&{\rm Im}\left([{\cal Z}]_{\alpha\beta}[{\cal Z}^*]_{\alpha\gamma}[{\cal Y}]_{\gamma\sigma}[{\cal Y}^*]_{\beta\sigma}\right)=|V_{4\alpha}|^2\,{\rm Im}\left([V_{4\beta}[V^*]_{4\gamma}[{\cal Y}]_{\gamma\sigma}[{\cal Y}^*]_{\beta\sigma}\right).
\ea
We inspected these structures and found that a non-vanishing correction to $\theta$ or $F_{u,d}$ in \eqref{FuFd} can only be obtained if subleading terms in the mixing $\sim Yv/M$ between $\psi$ and the SM are taken into account. As a result corrections to $\bar\theta$ due to loops of the Higgs boson are always smaller than in \eqref{nonAnalyWZest}.

\bibliography{biblio.bib}

\bibliographystyle{unsrturl}

\end{document}